\documentclass[aps,prl,twocolumn,amsmath,amssymb]{revtex4-1}
\usepackage{graphicx}
\usepackage{mathtools}
\usepackage{array}
\usepackage{xcolor}
\definecolor{green}{rgb}{0.0, 0.5, 0.13}
\DeclareGraphicsExtensions{.png,.eps,.pdf}

\DeclareMathOperator\arctanh{arctanh}
\begin{document}

\title{ Nearly order from quantum disorder phenomena and
its observation in the bosonic quantum anomalous Hall system }
{\sl  Substantially revised version of arXiv:1903.11134 }
\author{ Fadi Sun and Jinwu Ye  }
\affiliation{Department of Physics and Astronomy,
Mississippi State University, Mississippi State, Mississippi 39762, USA\\
Department of Physics, Capital Normal University,
Beijing 100048, China}

\date{\today }


\begin{abstract}
We report a new many body phenomena called " Nearly order from quantum disorder phenomena" (NOFQD).
We demonstrate the NOFQD  in the experimentally realized weakly interacting Quantum Anomalous Hall system
of spinor bosons in an optical lattice.
At a zero Zeeman field $ h=0 $, there is a classically infinite degeneracy
due to a spurious spin $ SU(2) $ symmetry.
By developing a systematic order from quantum disorder (OFQD) analysis,
we find the quantum ground state to be a $ N=2 $ XY-AFM superfluid (SF) state and also evaluate its excitation spectrum
due to the effective potential generated by the OFQD at $ h=0 $.
At a high $ h $, the system is in a Z-FM SF state.
At a small $ h>0 $, even there is no spurious symmetry anymore,
there is still a delicate competition between the Zeeman field effect at $ h >0 $ and
the effective potential
leading to a quantum phase transition between the two states.
We name this novel many body phenomena as NOFQD which describes how a OFQD state response non-trivially to a small deformation.
Starting from symmetry principle, we construct a Ginsburg-Landau (GL) type of effective action to
investigate the nature of the transition.
We establish intrinsic  connections between the phenomenological GL theory and the microscopic calculations on
the effective potential.
Connections with the bilayer quantum Hall system with a total filling factor $ \nu_T=1 $ are made.
Some insightful analogy with $ NAdS_2/NCFT_1 $ ( where $ N $ also means nearly ) correspondence in the context of Sachdev-Ye-Kitaev models are hinted.
Two types of OFQDs are classified, one response trivially, another non-trivially to a small deformation to the Hamiltonian leading to NOFQD.
The NOFQD can be detected in the current cold atom  bosonic quantum anomalous Hall experiments and may also appear
in many other frustrated systems.
\end{abstract}

\maketitle

{\bf 1. Introduction }

The investigation and control of spin-orbit coupling (SOC)
have become subjects of intensive research
in both condensed matter and cold atom systems
after the discovery of the topological insulators \cite{kane,zhang}.
In materials side, the SOC plays crucial roles
in various 2d or layered insulators, semi-conductor systems,
metals and superconductors without inversion symmetry.
The quantum anomalous Hall (QAH) effect was
experimentally realized in Cr doped Bi(Sb)$_2$Te$_3$ thin films
\cite{QAHthe,QAHexp}.
In the cold atom side, using Raman schemes,
several experimental groups \cite{expk40,expk40zeeman,2dsocbec}
generated 2d SOC for neutral cold atoms
in both continuum and in optical lattices.
Especially, the bosonic analog of the QAH
for spinor bosons $^{87}$Rb was realized in \cite{2dsocbec},
and the lifetime of SOC $^{87}$Rb Bose-Einstein condensation (BEC) have been improved from $300$ ms to $900$ ms recently.

Motivated by the recent experiment on bosonic QAH \cite{2dsocbec},
we investigate possible quantum many body phenomena in the bosonic quantum anomalous Hall (BQAH) model in a square lattice.
In contrast to its fermonic counterpart, the bosonic QAH system is necessarily interacting.
Most importantly, we discover a new many body phenomena we name " Nearly order from quantum disorder phenomena" (NOFQD).
Our main results are presented in Fig.1. We achieve Fig.1 by the combination of specific microscopic calculations and
the symmetry based phenomenological classification approaches.

We first perform microscopic calculations at  weak interaction limit and also identify its limitations.
At a high Zeeman field $ h $, we find the system is in a spin-polarized  SF state along the $ z $ direction ( named as Z-FM SF state ). We evaluate the SF Goldstone mode at $\vec{k}=(0,0) $ and the roton mode at $\vec{Q}=(\pi,\pi) $.
We delineate the behaviour of the roton mode as $ h $ decreases: the roton mode continues to drop
and eventually becomes a gapless quadratic mode $ \sim v^2k^2 $ as $ h \rightarrow 0^+ $.
At zero Zeeman field $ h=0 $, we identify a spurious spin $ SU(2) $ symmetry
which leads to  a classically infinitely degenerate ground state manifold.
We develop a systematic order from quantum disorder (OFQD) analysis to determine
an effective potential generated by the the OFQD effects. It selects the true quantum ground state from such
a manifold to be a $ N=2 $ XY-AFM SF state where $ N=2 $ stands for the number of BEC momenta.
It also opens a gap to the spurious quadratic roton mode.
However, at any small $ h >0 $, the spurious spin $ SU(2) $ symmetry disappears, a simple classical analysis leads to a unique classical ground state which is the Z-FM SF. Although  at a high $ h $, the system is indeed in this Z-FM SF state, the classical picture may break down
at a small $ h>0 $. At a small $ h >0 $, even there is not any spurious symmetries,
there is still a delicate competition between the classical  Z-FM SF state at a small $ h >0 $ and
the $ N=2 $ XY-AFM superfluid (SF) state selected by the OFQD mechanism at $ h=0 $
which may lead to a quantum phase transition between the two states.
We name this new quantum many body phenomena as " Nearly order from quantum disorder phenomena “ (NOFQD) which
happens slightly away from the conventional OFDQ in the parameter space of a Hamiltonian.
At the mean field level, we find that at a finite small $ h $, the competition between the Zeeman energy and the effective potential
first leads to a canted $ N=2 $ XY-AFM SF state (CAFM), then drives a transition to the Z-FM state.
Unfortunately, it becomes difficult to study the nature of the quantum phase transition
from the microscopic weak coupling  approach.

Then we change our strategy to symmetry based classification approach. Just from the constraints of the symmetries of the Hamiltonian,
we construct a phenomenological Ginsburg-Landau (GL) type of effective action to investigate
the NOFQD phenomena and the associated quantum phase transition from the a weak to strong Zeeman field.
Using the GL action, we first find the excitations in the CAFM and Z-FM at low and high Zeeman field respectively which include
the SF mode and the roton mode. It is the roton gap closing which drives the quantum phase transition from the CAFM to the Z-FM.
Then by performing a renormalization group (RG) analysis, we find the transition is in the same universality class as
the zero density Mott to superfluid transition,
subject to a marginally and dangerously irrelevant $ U(1) $ symmetry breaking  ( cubic ) anisotropic term.
So it has the critical exponents $ z=2, \nu=1/2, \eta=0 $ subject to new logarithmic corrections.
This $ U(1) $ symmetry breaking cubic anisotropic term leads to experimentally observable logarithmic corrections to
various scaling functions in the quantum critical regime.

By comparing our symmetry based GL action with the microscopic calculations at $ h=0 $ and a high $ h $, we establish the intrinsic and
profound connections between the phenomenological GL theory and the microscopic calculations, especially the
effective potential generated by the OFQD at $ h=0 $, therefore endow the broad impacts of OFQD from a new and profound  perspective.
The microscopic calculations and the phenomenological GL theory are two different, but complementary approaches.
Each has its own advantages and limitations. The combination of the two approaches lead to rather deep and complete understandings of
the NOFQD phenomena in the bosonic QAH systems.
We also make instructive mappings between the GL theory here and the Chern-Simon  GL theory
applied by one of the authors to study the exciton superfluid and quantum phase transitions (QPT)
in the bilayer quantum Hall systems at the total filling factor $ \nu_T=1 $.
The scaling behaviours of various physical quantities such as the roton gaps, the specific heat, transverse magnetization and
the Lyapunov exponent/butterfly velocity
are derived.

\begin{figure}
\includegraphics[width=\linewidth]{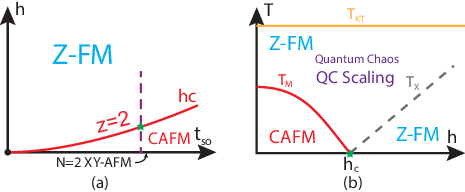}
\caption{ The quantum phase transition associated due to the "Nearly order from quantum disorder phenomena" NOFQD phenomena in the
interacting bosonic QAH system (a) Zero temperature phase diagram \cite{origin} as a function of
	$t_\text{so}/t_0$ and $h/t_0$ with a fixed $n_0U$. The CAFM SF phase is labeled by
$ \langle \psi_1 \rangle \neq 0,  \langle \psi_2 \rangle \neq 0 $, while
the Z-FM SF phase labeled by $ \langle \psi_1 \rangle \neq 0,  \langle \psi_2 \rangle = 0 $ or the other way around.
The quantum phase transition is in the same universality class as the zero-density SF to Mott transition subject to
the dangerously marginally irrelevant cubic anisotropic term described by Eq.\ref{critical2}.
	(b) Finite temperature phase diagram in the $ (h, T) $ plane with
	fixed $t_\text{so}/t_0$ and $n_0U$ ( along the dashed line in (a) ).
	Only the part below $ T < T_{KT} $ is shown. The dashed line $ T_X \sim \Delta_R $ scales as the roton gap
    and stands for the crossover from the QC regime to the gapped regime in the Z-FM SF phase. }
\label{QCPhase}
\end{figure}

In the conclusion section, in order to spell out the crucial connections and differences between the  OFQD and NOFQD. we classify the OFQD into two classes:
Type-I responses trivially to a deformation $ \lambda $,  type-II does non-trivially and leads to the NOFQD.
Some interesting and insightful analogy with $ NAdS_2/NCFT_1 $ ( where $ N $ also means nearly )
correspondence in the context of Sachdev-Ye-Kitaev models are also hinted.
The ongoing experimental efforts to detect these novel NOFQD phenomena are critically examined.
Some future perspectives are outlined.

The rest of the paper is organized as into 4 parts.
 In part I consisting of Sec.2-5, we perform microscopic calculations at $ h=0 $ and strong $ h \gg U $ at weak coupling limit.
 In part II consisting of Sec.6-9, to capture the NOFQD, we perform phenomenological effective action calculations at any $ h $ at weak coupling limit.
 Then in Part III consisting of Sec.10-11, we contrast the phenomenological effective action calculations in part II with the microscopic calculations in part I, therefore establish the intrinsic connections between the two complementary approaches.
 We discuss the experimental implications and  reach conclusions in Sec.12 and 13 respectively.


{\bf 2.
The classical ground state at a high field $h \gg U$ and its possible breakdown in the small $h \lesssim U$ limit.}

The experimentally realized quantum anomalous Hall model
of spinor bosons in a square lattice is described by
the Hamiltonian \cite{2dsocbec}
\begin{align}
    \mathcal{H}=
	&-t_0\sum_{\langle ij\rangle}
	a_{i}^\dagger \sigma^z a_{j}
	+it_\text{so}\sum_{\langle ij\rangle}
	a_i^\dagger(\mathbf{d}_{ij}\cdot\vec{\sigma}) a_{j} \nonumber\\
	&-h\sum_i a_{i}^\dagger\sigma^z a_{i}
	+\frac{U}{2}\sum_i n_i(n_i-1)
	-\mu\sum_i n_i
\label{eq:model}
\end{align}
where 
$a_i=(a_{i\uparrow}, a_{i\downarrow})^\intercal$
and $\langle ij\rangle$ denotes a pair of nearest neighbor sites,
and $\mathbf{d}_{ij}=\mathbf{r}_i-\mathbf{r}_j$.
$U$ is the repulsive onsite interaction,
and $\mu$ is the chemical potential,
$\sigma_{x,y,z}$ are three Pauli matrices,
and $n_i=a_i^\dagger a_i$ is the number of particles on site-$i$.
For the $^{87}$Rb atoms used in the recent experiment \cite{2dsocbec},
the two pseudo-spin components $\sigma=\uparrow,\downarrow$
denote the two hyperfine states $|F=1,m_F=0\rangle$ and $|F=1,m_F=-1\rangle$.

The Hamiltonian has the particle number conversation U$_c$(1) symmetry,
and a spin-orbital coupled $[C_4\times C_4]_D$ symmetry.
Under the mirror transformation
$ {\cal M} =(-1)^{i} R_z(\pi) {\cal T} $, $ h \rightarrow -h $, so $ h=0 $ has an enlarged Mirror symmetry.
So we only need to focus on $h\geq0$ cases.
In this Letter,  we focus on the experimentally accessible regime $0\leq t_\text{so}\leq t_0$ and $U/t_0\ll1$.
The global phase diagram in most general parameter space
will be worked out in a much longer version \cite{our}.

Any small $ h > 0 $ breaks the mirror symmetry and
split the two degenerate single-particle states
at momentum $(0,0)$ and $(\pi,\pi)$.
A straightforward mean field analysis predicts the condensation occurs at
either $(0,0)$ or $(\pi,\pi)$ depending on the sign of $h$.
Due to its spin alinement along the $ z $ axis, we name it Z-FM superfluid in Fig.1.

 When the spectrum minimum is located at $(0,0)$ or $(\pi,\pi)$, the non-interacting Hamiltonian takes a simple form
\begin{align}
    H_0(k)=-[h+2t_0(\cos k_x+\cos k_y)]\sigma_z
\end{align}
   where the SOC $ t_\text{so} $ drops out.

   Let us consider $h>0$ case, thus the minimum is located at $(0,0)$.
In the weak coupling limit $ U/t \ll 1 $, by writing
\begin{align}
	a_{k\uparrow}\to\sqrt{N_0}\delta_{k,0}+\psi_{k\uparrow},\quad
	a_{k\downarrow}\to\psi_{k\downarrow}
\end{align}
  where $N_0$ is the number of condensate atoms,
  and $N_s$ is the total number of lattice sites, thus $n_0=N_0/N_s$ is the condensate fraction.

   We can perform the expansion $  \mathcal{H}=\mathcal{H}^{(0)}+\mathcal{H}^{(1)}+\mathcal{H}^{(2)}+\cdots
    $ where the superscript denotes the order in the quantum fluctuations.
    The zeroth order term $ \mathcal{H}^{(0)}=E_0= -\frac{1}{2}U n_0N_0 $  is the classical energy of the condensate.
  The vanishing of the linear term sets the value of the chemical potential $\mu=-h-4t_0+U n_0$.
  Diagonizing  $ \mathcal{H}^{(2)} $ by a generalized $ 4 \times 4 $ Bogliubov transformation leads to:
\begin{align}
     H=E_{0}+ E^{(2)}_0 + \sum_{k,s=\pm} \omega_s(k)( \alpha_{sk}^\dagger\alpha_{sk} + 1/2)
\label{e0e2modes}
\end{align}
where  $ E^{(2)}_0=-( h+4t_0+ \frac{1}{2} U n_0 ) $
and $\omega_{\pm}(k)$ are the two Bogliubov modes.

\begin{figure}[!htb]
\centering
\includegraphics[width=0.65\linewidth]{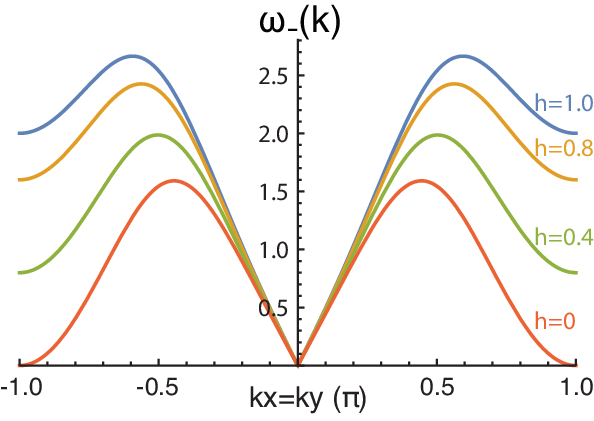}
\caption{
	The Critical Behaviour of the superfluid Goldstone mode $\omega_-(k_x=k_y)$
	starting at $ ( t_\text{so}/t_0=1, h/t_0=1 ) $.
Decreasing $h/t$ from 1 to 0, the roton at $ \vec{Q}= (\pi, \pi) $ gets lower and lower and drops to zero at $ h=0 $
signifying a possible ( in fact, naive) first order transition.}
\label{roton1}
\end{figure}

Since $\omega_+(\mathbf{k})>\omega_-(\mathbf{k})$ always holds, so we focus on $\omega_-(\mathbf{k})$
which displays a gapless superfluid Goldstone mode near $ \mathbf{k}=(0,0) $ and
a gapped roton mode at $ \mathbf{k}=(\pi,\pi) $. In the long wavelength limit, we find
the superfluid Goldstone mode becomes linear
$\omega_-(k)=c|\mathbf{k}| $ where the SF velocity is
\begin{align}
    c=\sqrt{\frac{2(4t_0^2-2t_\text{so}^2+th)Un_0}{4t_0+h}}
\label{slope}
\end{align}
  which is shown in Fig.\ref{roton1}a. In the $ h \to 0 $ limit, it reduces to $ c= \sqrt{ n_0 U t (2- t^{2}_s/t^2) } $.


In Fig.\ref{roton1}a, at a fixed $ t_\text{so}=1 $, as $ h $ decreases to $ 0 $ sitting on the boundary between  Z $ \uparrow $ FM-SF at $ (0,0) $ and
  the Z $ \downarrow $ FM-SF at $ (\pi,\pi) $, the roton mode gets lower and lower,
  then touches zero at $ \vec{Q}= (\pi,\pi) $. In the $ h \rightarrow 0 $ limit, the roton mode can be
  extracted from $\omega_-(k)$ mode near $k=(\pi,\pi)+q$:
\begin{align}
	\omega_-(k)
	=\sqrt{4h^2+v_{4}(q_x^2+q_y^2)^2}
\end{align}
where
\begin{align}
	v_{4}=\frac{(2t_0^2-t_\text{so}^2)[t_0(4t_0+n_0U)-2t_\text{so}^2]}{2t_0(4t_0+n_0U)}
\end{align}
which depends on the the interaction $ n_0 U $.

In the weak coupling limit $n_0U\ll t_0$, the $v_{4}$ can be simplified as
$ v_{4}=(2t_0^2-t_\text{so}^2)^2/(4t_0^2)=(v^2)^2 $ which remains finite even at the non-interacting limit $ n_0 U=0 $.
In the small $ h $ limit, the roton gap
\begin{equation}
 \Delta^0_R(h)=2|h|
 \label{rotonh}
\end{equation}
 which is independent of $ t_\text{so} $ and shown in the dashed line in Fig.4b.

As one decreases $h$ from $ h \gg U $ to $ h \lesssim U $, then to $0^{+}$,
the roton mode gets lower and lower, then touches zero at $ h=0^{+} $,
it reduces to the quadratic form $ \omega_-(k)=\sqrt{v_{4}(q_x^2+q_y^2)^2}=v^2 q^2$.
This behaviour may signify a possible first order transition at $h_c=0^{+}$.
This physical picture leads to the naive phase diagram shown in Fig.S1 in SM.
However, as to be shown below, this naive physical picture only holds when $ h $ is sufficiently large $ h > U $,
but may break down \cite{bert} at a small $ h \lesssim U $.

{\bf 3. The order from quantum disorder (OFQD) effects to determine the quantum ground state at $ h=0 $.}

The single-particle energy spectrum $ U=0 $  of Eq.\eqref{eq:model} in the momentum space can be easily obtained:
$\epsilon_{\pm,\mathbf{k}}^h
=\pm\sqrt{[h+2t_0(\cos k_x+\cos k_y)]^2
+4t_\text{so}^2(\sin^2 k_x+\sin^2 k_y)}$.
At $h=0$, its lower branch develops two degenerate minima at
momentum $(0,0)$ and $(\pi,\pi)$,
and two eigen-spinors are $\eta_0=(1,0)^\intercal$
and $\eta_\pi=(0,1)^\intercal$ respectively.
There is a classically degenerate family of states:
\begin{equation}
  \Psi_i=c_0\eta_0+(-1)^{i_x+i_y} c_\pi \eta_\pi
\label{deg}
\end{equation}
where the two complex number $c_0$ and $c_\pi$
satisfy normalization condition $|c_0|^2+|c_\pi|^2=1$.
This classically degenerate manifold is due to the spurious $ SU(2)_s $ symmetry at the mean field level.

By writing the spinor field as
$ \Psi= \sqrt{N_0} \Psi_0+ \psi $, we can expand the Hamiltonian $ \mathcal{H}=\mathcal{H}^{(0)} + \mathcal{H}^{(1)}+\mathcal{H}^{(2)} +\cdots $.
The zeroth order term $ \mathcal{H}^{(0)} $ gives the classical ground state
energy $ E_0=-\frac{1}{2}Un_0N_0 $.
Setting the linear term $ \mathcal{H}^{(1)} $ vanish gives the value of the chemical potential $\mu=-4t+Un_0$.
Diagonalizing the quadratic term  $ \mathcal{H}^{(2)} $ by a generalized $ 8 \times 8 $  Bogliubov transformation \cite{RBZ} leads to:
\begin{align}
    \mathcal{H}=E_0+E_\text{ofd}
	+\sum_{l,\mathbf{k}\in RBZ} \omega_{l,\mathbf{k}}
	\alpha_{l,\mathbf{k} }^\dagger\alpha_{l,\mathbf{k}}
\label{ofdleft}
\end{align}
where $E_\text{ofd}=-(4t_0+\frac{1}{2}n_0U)N_s
+\frac{1}{2}\sum \omega_{l,\mathbf{k}}$
is quantum correction to the mean field ground-state energy,
and $\omega_{l,\mathbf{k}}$ with $l=1,..,4$ represent 4 Bogoliubov spectrum.

It is convenient to parameterize $c_0$ and $c_\pi$ as
\begin{align}
    c_0=e^{-i\phi/2}\cos(\theta/2),\quad
    c_\pi=e^{i\phi/2}\sin(\theta/2)
\label{eq:theta_phi}
\end{align}
then we can plot numerically  $E_\text{ofd}(\theta,\phi)$
as a function of $\theta$ and $\phi$ in Fig.2 where one can identify
the quantum ground states as $\theta=\pi/2$ and $\phi_m=\pi/4+m\pi/2$ ($m=0,1,2,3$).
It  has a uniform density $\langle n_i\rangle=n_0$ and
a XY-AFM ordered spin structure: $\langle a_i^\dagger\vec{\sigma} a_i\rangle
=(-1)^{i_x+i_y}n_0(\cos\phi_m,\sin\phi_m,0)$. It is 4 fold degenerate breaking the  joint $ C_4 $ symmetry.

\begin{figure}[!htb]
\includegraphics[width=\linewidth]{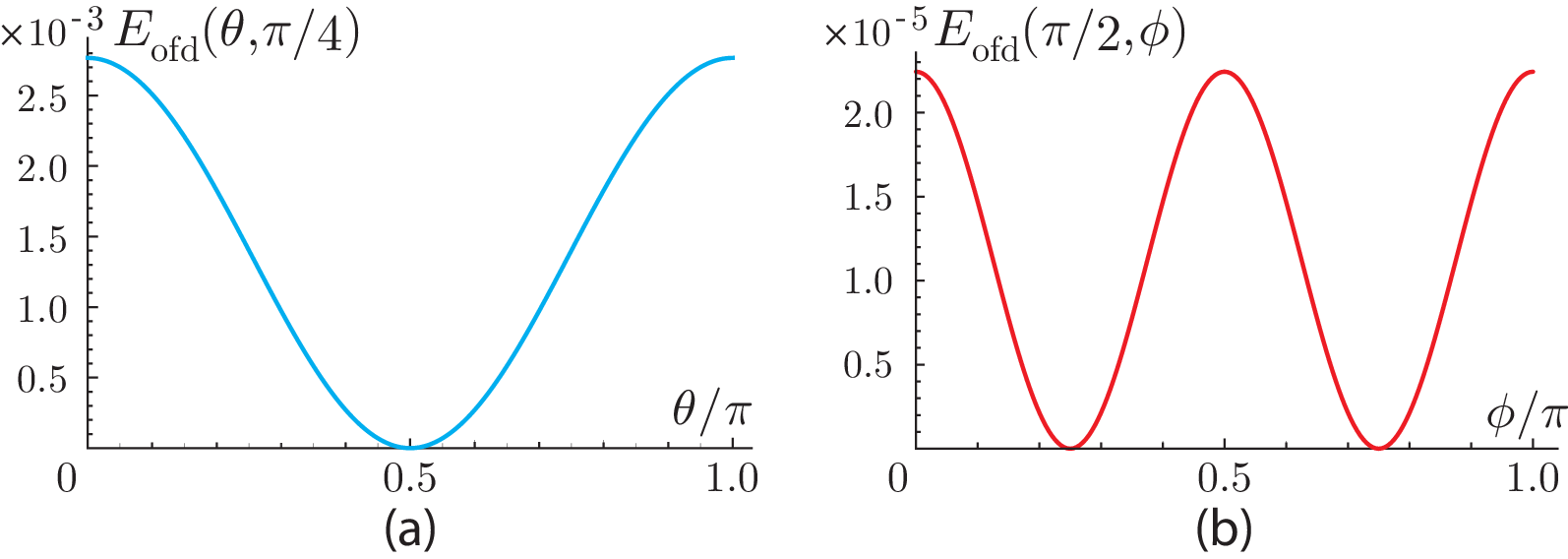}
\caption{The quantum ground-state energy density near its minimum
	at (a) $\theta=\pi/2$ at fixed $\phi=\pi/4$,
	(b) $\phi=\pi/4,3\pi/4$ at fixed $\theta=\pi/2$
	where the coefficients $A$ and $B$ can be extracted respectively.
	We used $n_0U/t_0=1$ and $t_\text{so}/t_0=1$. Despite both $ A $ and $ B $ are generated by the OFQD, the numerical value of $ A $
    is larger than $ B $ by 2 orders of magnitude.}
\end{figure}

After identifying the quantum ground-state as the $ N=2 $ XY-AFM SF state,
we can also evaluate all $\omega_{1,2,3,4}(k)$ in Eq.\ref{ofdleft}.
There are one linear $\omega_1 \sim k $ SF Goldstone mode and
one $\omega_2 \sim k^2 $ quadratic roton mode located at $(0,0)$.
While $\omega_3(k)$ and $\omega_4(k)$ are two fully gapped higher energy modes.


{\bf 4. The Roton mass gap generated from the OFQD mechanism in the $ N=2 $ XY-AFM state at $ h=0 $. }

Now we can have a better understanding of the spurious $ SU(2)_s $ symmetry and the origin of this spurious qudratic roton mode:
Indeed, it setting the two minima $ (0,0) $ and $ (\pi,\pi) $ into the kinetic energy, then the term involving $ t_{so} $ drops out,
so at the mean field level, the system has the $ SU(2)_s $ symmetry originating from the exact $ SU(2)_s $ symmetry at $ t_{so}=0 $.
While the $ N=2 $ XY-AFM state is a collinear state invariant under the rotation
$ e^{i \alpha \vec{\sigma} \cdot \vec{n} }, \vec{n}=(1,1,0) $, in fact, it is a XY-FM state in the rotated $ \tilde{SU}(2) $ basis
$\tilde{\mathbf{S}}_i=((-1)^{i_x+i_y}S_i^x,(-1)^{i_x+i_y}S_i^y,S_i^z)$ where the  $ SU(2)_s $ symmetry
is manifest ( See appendix B ). So its breaking to the FM state leads to the spurious quadratic FM roton mode which is remanent of the exact FM mode at $ t_{so}=0 $. However, the OFQD opens a gap to this spurious quadratic FM roton mode.

The OFQD mechanism picks up $\theta=\pi/2$ and $\phi=\pi/4$ as the quantum ground state.
Then we can expand the ground-state energy around the minimum as \cite{more}:
\begin{align}
	E_\text{GS}[\theta,\phi]=E_0+\frac{A}{2}\delta\theta^2+\frac{B}{2}\delta\phi^2
\label{AB}
\end{align}
where $\theta=\pi/2+\delta\theta$ and $\phi=\pi/4+\delta\phi$, the two coefficients $ A \sim (n_0 U)^2/t $ and
$ B \sim (n_0 U)^4/t $ can be extracted from Fig.M2.

  Using the commutation relations $ [\frac{1}{2}n_0\delta \theta, \delta \phi ]= i \hbar $, one can see that
  the "quantum order from dis-order " mechanism generates a roton gap:
\begin{equation}
  \Delta_R( h=0 )=2\sqrt{AB}/n_0
\label{roton0}
\end{equation}
  which is the dashed line in Fig.M3b  ( also in the Fig.S1b in SM  ).
  The SF Goldstone mode remains unaffected due to its protection by the $ U(1) $ symmetry breaking.

{\bf 5. The nearly order from quantum disorder (NOFQD) effects  at a small $ 0< h \lesssim U $:
 Mean field theory treatment in the microscopic calculation }

Any small $ h > 0 $ spoils the spurious $ SU(2)_s $ symmetry and also
splits the degenerate single-particle state
at momentum $(0,0)$ and $(\pi,\pi)$. So one can determine the ground state to be the classical Z-FM SF state
even at an infinitesimal small $ h $.
However, when $ h \lesssim U $, the splitting is so small that any small interaction $ U $
still mix the two states considerably. So even the spurious $ SU(2)_s $ symmetry disappears at any small $ h $,
one must still take into account the OFQD effects at $ h=0 $
and study the delicate  competition between the OFQD generated effective potential in Fig.2 due to the
interaction $ U $  with the Zeeman energy due to a small finite $ h $.
We call this phenomena as NOFQD.

As shown in SM, the OFQD generated effective potential in Fig.2 can be cast into the form:
\begin{align}
    E_\text{ofd}
	=& E_\text{ofd,0}+N_s\Big[\frac{1}{4}A(1+\cos2\theta)\nonumber\\
	&+\frac{1}{64}B(1-\cos2\theta)^2(1+\cos4\phi)+\cdots\Big].
\label{eq:Eofd}
\end{align}
  Appealingly, one can cast the energy density:
$ \mathcal{E}_\text{ofd}=E_\text{ofd}/N_s$ into a compact from
\begin{align}
   \mathcal{E}_\text{ofd} 
	=
	\frac{1}{2}A  (|c_0|^2\!-|c_\pi|^2)^2
	+\frac{1}{2}B  [(c_0c_\pi^*)^2\!+(c_0^*c_\pi)^2]^2
\label{eq:Ec0cpi}
\end{align}
where the coefficients $A$ and $B$ are positive numbers at least
in the weak coupling limit.



To get to finite temperatures, it is necessary to parameterize the quantum fluctuations in
the polar coordinate as:
\begin{align}
    a_{i}(\tau)
	=\sqrt{n_i(\tau)}
	e^{i\chi_i(\tau)}
	[c_{0,i}(\tau) \eta_{0}
	+(-1)^ic_{\pi,i}(\tau) \eta_{\pi}]
\label{deg2}
\end{align}
   and write the action as two parts
$\mathcal{S}=\int d\tau(\mathcal{L}+\delta\mathcal{L}_c)$.
\begin{eqnarray}
    \mathcal{L}
	& = & \sum_\mathbf{k}\bar{a}_\mathbf{k}
	[\partial_\tau+\epsilon_{-,\mathbf{k}}^{h=0}\!-\!\mu]a_\mathbf{k}
	+\frac{U}{2}\sum_in_i(n_i-1)            \nonumber   \\
    \delta\mathcal{L}_c
	& = & \sum_i [\mathcal{E}_\text{ofd}(c_{0,i},c_{\pi,i})-nh(|c_{0,i}|^2-|c_{\pi,i}|^2)]
\end{eqnarray}
where  the second part contains the competition between the
order-from-disorder effects generated effective potential and the Zeeman effect

  With the parametrization Eq.\eqref{eq:theta_phi}, we obtain
\begin{align}
    \mathcal{L}_0
	    =E_0
	    +\frac{A}{2}\cos^2\theta
	    +\frac{B}{8}\sin^4\theta\cos^22\phi
	    -n_0 h\cos\theta
\label{eq:newL}
\end{align}
where $E_0$ is independent of $\theta$ and $\phi$.
The $\cos4\phi$ term is a result of $C_4$ symmetry
and plays a similar role as the $C_4$ clock term.
Performing a minimization with respect to $\theta$ and $\phi$ leads to $\theta_0=\arccos(h/h_c)$ and $\phi_0=\pi/4$ with $h_c=A/n_0$.
In the $h\to 0^{+}$ limit, one recovers the $ N=2 $ XY-AFM SF state.
For $0<h<h_c$, the spin-orbital  structure is
$n_0((-1)^{i_x+i_y}\sin\theta_0\cos\phi_0,
(-1)^{i_x+i_y}\sin\theta_0\cos\phi_0,\cos\theta_0)$,
and named as a canted AFM (CAFM) SF state.
When $h \leq h_c$, the spin structure becomes $n_0(0,0,1)$ which is fully aligned to the $z$ direction, so named the $Z$-FM state (Fig.3a).

\begin{figure}[!htb]
\includegraphics[width=\linewidth]{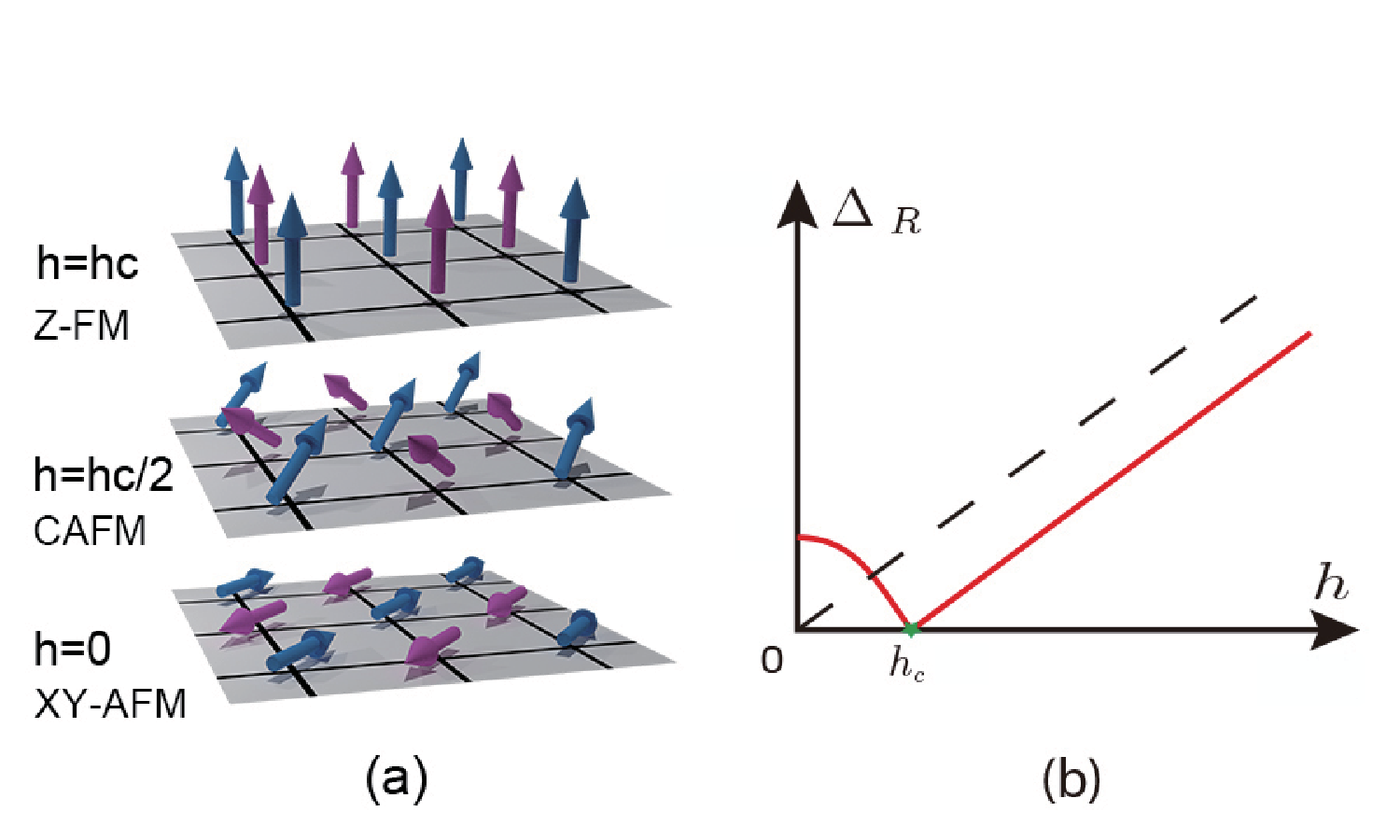}
\caption{(a) The spin-bond structure of the ground-state
	as the Zeeman field $h$ from $0$ to $h_c$ and beyond.
	(b) The Roton gap $\Delta_R$ from the second equation in Eq.\ref{lowpm} below $ h_c $ and Eq.\ref{highpm} above $ h_c $
     through a 2nd QPT.	The dashed line is the roton gap  Eq.\ref{rotonh} calculated by the naive Bogliubov calculation.}
\end{figure}

   Unfortunately, in this microscopic NOFQD calculation, the BEC condensation density $ n_0 $ is taken as a constant
   at very beginning ( see above  Eq.\ref{ofdleft} ). The chemical potential $ \mu=-h-4t + U n_0 $ also depends on $ n_0 $ and $ h $.
   It is very difficult to capture the density fluctuations in a consistent way.
   To study the competition between the effective potential generated by the OFQD and the Zeeman field, then
   to capture the nature of the quantum phase transition near $ h \sim h_c $, one must take account the density fluctuations.
   To circumvent this difficult suffered in this this microscopic NOFQD calculation,
   one may turn to an independent phenomenological calculations based on the symmetries of the Hamiltonian.


{\bf 6. Ginsburg-Landau type action from the symmetry principle to study the NOFQD phenomena: }

   To capture the density fluctuations in a consistent way, it is convenient to
   absorb the density/phase part into $ c_0 $ and $ c_{\pi} $ in Eq.\ref{deg2} and
   re-parameterize low energy quantum fluctuations at a small $ h \ll U $ as
\begin{align}
	a_{i\alpha}=\psi_1\eta_0+(-1)^i\psi_2\eta_\pi
\label{boson12}
\end{align}
   which includes both spin and charge fluctuations.

   Then one can find the density  $ n= |\psi_1|^2+|\psi_2|^2 $ and the spin:
\begin{align}
	S^{+} & =(-1)^i\psi_1^*\psi_2,   \nonumber  \\
    S^{-} &=(-1)^i \psi_2^*\psi_1,   \nonumber  \\
    S^{z} & =|\psi_1|^2-|\psi_2|^2
\label{spincom}
\end{align}

 Note that here we only expand the boson field in terms of the two minima of the kinetic term without assuming any symmetry breaking.
 In order to study the transition from the weak coupling SF to the strong
 coupling Mott phase an an integer filling, we construct the Ginsburg-Landau action consistent with all the symmetries
 of the Hamiltonain such as the translational symmetry, the $ U(1)_c $ symmetry, the $ [ C_4 \times C_4 ]_D $
 ( or $ [R_{\pi/2} \times R_{\pi/2}]_D $ ) spin-orbital coupled symmetry
 and the  Mirror symmetry $ {\cal M} $ at $ h=0 $:
\begin{align}
    \mathcal{L}
	& = \psi_1^*\partial_\tau\psi_1+\psi_2^*\partial_\tau\psi_2
	+v^2(|\nabla\psi_1|^2+|\nabla\psi_2|^2)    \nonumber\\
    &+r(|\psi_1|^2+|\psi_2|^2) +\frac{U}{2}(|\psi_1|^2+|\psi_2|^2)^2   \nonumber\\
	& +\frac{A}{2}(|\psi_1|^2-|\psi_2|^2)^2
	+\frac{B}{2}[(\psi_1^*\psi_2)^2+(\psi_1\psi_2^*)^2]^2     \nonumber\\
	& -h(|\psi_1|^2-|\psi_2|^2) + \cdots
\label{blqh}
\end{align}
where the $ \cdots $ means the higher order terms consistent with all the symmetries ( see SM Eq.S25 ),
$r<0$ ensures the system is always in a superfluid phase,
and $v,U $ are all positive to make the system stable.
However, the sign of $ A $ and $ B $ can not be determined by the symmetry analysis, but can be
fixed by comparing with the microscopic calculations at the weak coupling limit.
The last $ h $ term is the Zeeman field coupled to $ S^z$ in Eq.\ref{spincom}.
It breaks the Mirror symmetry $ {\cal M} $.
The relations between these phenomenological parameters and the microscopic ones calculated in the previous sections
will be established in Sec.9.

Although this effective action was designed to study the transition from weak coupling to the strong coupling.
Here, we only use it to study the transition driven by the Zeeman field,
while leave the original goal to a future study \cite{strongcoupling}.

 It is convenient to write the two bosons in terms of its magnitude and phase
$\psi_\alpha=\sqrt{\rho_\alpha}e^{i\theta_\alpha}$,
then
\begin{align}
	\Omega & =r(\rho_1+\rho_2)+\frac{U}{2}(\rho_1+\rho_2)^2
	+\frac{A}{2}(\rho_1-\rho_2)^2              \nonumber\\
    &+ 2B\rho_1^2\rho_2^2\cos^2[2(\theta_1-\theta_2)]
	-h(\rho_1-\rho_2)
\end{align}
which leads to two sets of saddle point solutions.

For $ h< h_c=-rA/U=\rho_0 A $:
\begin{align}
	&\bar{\rho}_1  =(\rho_0+h/A)/2,\quad \bar{\rho}_2=(\rho_0-h/A)/2,    \nonumber \\
	&\theta_1-\theta_2  =\pi/4+m\pi/2, \quad m=0,1,2,3
\label{canted}
\end{align}
 where $ \rho_0=\bar{\rho}_1+\bar{\rho}_2= -r/U$ is the total density of the spinor atoms \cite{anotherwriting}.
 Note that at these saddle points, the effective mass term $ r= - \rho_0 U $ is independent of $ h $.
 From Eq.\ref{spincom}, $  S^z=\bar{\rho}_1-\bar{\rho}_2=h/A=\rho_0 \cos \theta_0 $, namely, $ \cos \theta_0=h/h_c $.
 So the saddle point Eq.\ref{canted} has the spin-orbital  structure
 $\rho_0((-1)^{i_x+i_y}\sin\theta_0,(-1)^{i_x+i_y}\sin\theta_0,\cos\theta_0)$ named as a canted AFM (CAFM) SF state (Fig.3a).
 In the $h\to 0^{+}$ limit, it reduces to the the $ N=2 $ XY-AFM SF state with $ \theta_0=\pi/2 $.
 When $h \rightarrow h^{-}_c$, the spin structure becomes $n_0(0,0,1)$ which is fully aligned to the $z$ direction, so named the $Z$-FM state (Fig.3a).

 For $ h > h_c $:
\begin{align}
\bar{\rho}_1=\rho_0,\quad \rho_2=0
\label{zfmsad}
\end{align}
  which, according to Eq.\ref{spincom}, is the Z-FM state with $ \theta_0=0 $ in Fig.3a.
  Note that at this saddle points, the effective mass term $r=h-\rho_0 (A+U)=(h-h_c)-\rho_0 U$ depends on  $ h $
  and $\lim\limits_{h\to h_c^+}r=-\rho_0 U$ which approaches to the same value as that at $ h \leq h_c $.

  At the mean field level, we reproduced the two different ground states  $ h< h_c $ and $ h>h_c $ respectively achieved by the microscopic approach used in the previous sections. In the following, we will explore the quantum fluctuations, especially the Quantum phase transition between the two.


{\bf 7. The canted anti-ferromagnetic (CAFM) SF phase at $h<h_c$. }

Below the critical point $h<h_c$, the bosonic system is in canted anti-ferromagnetic superfluid phase Eq.\ref{canted}.
Without of loss generality, we choose $\bar{\theta}_1=0$ and $\bar{\theta}_2=\pi/4$
and write fluctuations as
$\psi_\alpha=\sqrt{\bar{\rho}_\alpha+\delta\rho_\alpha}e^{i(\bar{\theta}_\alpha+\delta\theta_\alpha)}$,
for the notation simplicity, we drop the $ \delta $ in $ \delta\theta_\alpha $,
then
\begin{align}
    \mathcal{L}_\text{CA}
	&=i\delta\rho_1\partial_\tau \theta_1+i\delta\rho_2\partial_\tau \theta_2
	+v^2\bar{\rho}_1(\nabla \theta_1)^2 +v^2\bar{\rho}_2(\nabla \theta_2)^2     \nonumber\\
	&+\frac{U}{2}(\delta\rho_1\!+\!\delta\rho_2)^2
	+\frac{A}{2}(\delta\rho_1\!-\!\delta\rho_2)^2
	+8B\bar{\rho}_1^2\bar{\rho}_2^2(\theta_1\!-\!\theta_2)^2   \nonumber\\
    & +\frac{v^2}{4\bar{\rho}_1}(\nabla\delta\rho_1)^2+\frac{v^2}{4\bar{\rho}_2}(\nabla\delta\rho_2)^2
\end{align}
 where we still keep the gradient term on the density fluctuations
 in the 3rd line which are higher orders to the mass terms
 of the density fluctuations in the second line \cite{blqhdrop}.

 Following bilayer quantum Hall systems \cite{blqh1,blqh2,blqh3},
 we introduce $ \bar{\rho}_\pm=\bar{\rho}_{1}\pm \bar{\rho}_{2},
 \delta \rho_\pm= \delta \rho_{1}\pm \delta\rho_{2}$ and  $\theta_\pm=\theta_{1}\pm\theta_{2}$, then
\begin{align}
    \mathcal{L}_\text{CA}
	&=\frac{i}{2}\delta\rho_+\partial_\tau \theta_+
	\!+\!\frac{i}{2}\delta\rho_-\partial_\tau \theta_-
	\!+\!\frac{v^2 \rho_0}{4}[(\nabla \theta_+)^2\!+\!(\nabla \theta_-)^2]  \nonumber\\
    &+\frac{U}{2}(\delta\rho_+)^2
	+\frac{A}{2}(\delta\rho_-)^2
	+\frac{B}{2}(\rho^2_0-\bar{\rho}_-^2)^2( \theta_-)^2     \nonumber\\
    & +\frac{v^2\rho_0}{4(\rho^2_0-\bar{\rho}_-^2)}
	[(\nabla\delta\rho_+)^2+(\nabla\delta\rho_-)^2]                     \nonumber\\
    & +\frac{v^2\bar{\rho}_-}{2}\nabla \theta_+\!\cdot\!\nabla \theta_-
    -\frac{v^2\bar{\rho}_-}{2(\rho^2_0-\bar{\rho}_-^2)}
	\nabla\delta\rho_+\!\cdot\!\nabla\delta\rho_-
\label{actionpm}
\end{align}
  where the $ B $ terms leads to the mass term of $ \theta_- $, the last line
  is the coupling between the $ + $ and $ - $ mode \cite{blqhdrop}.
  In the absence of the Zeeman field $ h=0 $, $ \bar{\rho}_-=0 $, then the $ + $ and $ - $ modes decouple,
  this corresponds to the balanced BQHE.

  The general expressions for the two eigen-modes are complicated. So we only list its long-wavelength limit:
\begin{align}
	\omega^{+}_\text{CA}(k) &=\sqrt{2\rho_0Uv^2k^2}  \nonumber  \\
	\omega^{-}_\text{CA}(k) &=\sqrt{ \Delta^2_R + v^2_R k^2 }
\label{lowpm}
\end{align}
where $\rho_0=\bar{\rho}_1+\bar{\rho}_2=-r/U, \theta_0=\arccos(h/h_c)$
and
\begin{eqnarray}
\Delta_R  & =  & 2\sqrt{B/A^3}(h^2_c-h^2),   \nonumber  \\
 v^2_R  &  = &  2\rho_0(A+B\rho_0\sin^2\theta_0)v^2
 \label{RR}
\end{eqnarray}
The $ + $ mode is nothing but the superfluid Goldstone mode.
While the $ - $ mode is the roton mode whose energy takes a relativistic form with a roton gap:
 vanishes  as $ \Delta_R \sim h_c-h $ when $ h \rightarrow h- h^{-}_c $.
 Obviously, it is the $ B $ term which opens the gap to the roton mode.

{\bf 8.  The Z-FM SF phase at  $h \geq h_c$.}

Above the critical point $ h > h_c $,
it is more convenient to introduce a mixed coordinate system
with $\psi_1=\sqrt{\bar{\rho}_1+\delta\rho_1}e^{i \theta_1 }$   in the polar coordinate and $ \psi_2 $
in the Cardisian coordinate. So $ (\delta\rho_1, \theta_1 ) $ and $\psi_2$ are all small, the Lagrangian density becomes
\begin{align}
    \mathcal{L}_\text{FM}
	 &=i\delta\rho_1\partial_\tau \theta_1
	+v^2 \rho_0(\nabla \theta_1)^2
	\!+\frac{m^2_1}{2}(\delta\rho_1)^2 \!+\frac{v^2}{4\rho_0}(\nabla\delta\rho_1)^2 \nonumber\\
	&+\psi_2^*\partial_\tau\psi_2
	+v^2|\nabla\psi_2|^2+r_2|\psi_2|^2
	+\frac{1}{2}(U+A)|\psi_2|^4    \nonumber\\
	&+(U-A)\delta\rho_1|\psi_2|^2 +B(\rho_0+\delta\rho_1)^2|\psi_2|^4  \nonumber\\
	&+\frac{B}{2}(\rho_0+\delta\rho_1)^2
	[(e^{-i\theta_1}\psi_2)^4+(e^{i \theta_1}\psi_2^*)^4]
\end{align}
where $  m^2_1=U+A 4 $ and $r_2=r+h+(U-A)\rho_0=2(h-A\rho_0)=2(h-h_c)$.

 To study the universality class of the transition, it is sufficient  only to keep
 the lowest order coupling terms:
\begin{align}
    \mathcal{L}_\text{FM}
	&=\psi_2^*\partial_\tau\psi_2+v^2|\nabla\psi_2|^2
	+r_2|\psi_2|^2+u_2 |\psi_2|^4    \nonumber\\
    &+ i \Gamma \delta\rho_1\partial_\tau\theta_1
	+v^2\bar{\rho}_1(\nabla\theta_1)^2
	+\frac{1}{2}m^2_1(\delta\rho_1)^2	  \nonumber\\
    &+ \gamma_0 \delta\rho_1|\psi_2|^2 +\gamma_1
	[e^{-i4 \theta_1}\psi^4_2+e^{i4 \theta_1}\psi^{*4}_2 ]
\label{critical}	
\end{align}
  where we dropped $ (\nabla\delta\rho_1)^2 $ term which is irrelevant anyway \cite{blqhdrop},
  $  u_2=\frac{1}{2}(U+A+ 2 B\bar{\rho}^2_1 ), \gamma_0=U-A, \gamma_1=\bar{\rho}^2_1 B/2 $.
  $ \Gamma=1 $ at the bare level, but was introduced to keep track of its flow under RG.
  The first line is the quantum critical mode of $ \psi_2 $ in the universality class of zero density
  superfluid to Mott (SF-Mott) transition, therefore has the critical exponents $ z=2, \nu=1/2, \eta=0 $.
  The second line stands for the gapless SF mode of $ (\delta\rho_1, \theta_1 ) $.
  The third line is the coupling between the two sectors with the coupling parameters $ \gamma_0,\gamma_1 $.

  From the quadratic parts of  Eq.\ref{critical}, one can

   extract the two eigen-modes
  in the long wavelength limit:
\begin{align}
	\omega^+_\text{FM}&=\sqrt{2\rho_0(A+U)v^2k^2}    \nonumber  \\
	\omega^-_\text{FM}&=2(h-h_c)+v^2k^2
\label{highpm}
\end{align}
which has the dynamic exponent $ z=2 $.

   Contrasting Eq.\ref{lowpm} with Eq.\ref{highpm}, one can see that
   due to the extra $ A $ term in the latter, the SF velocity increases  from $ h \ll h_c $ to $ h \gg  h_c $,
   while the difference $ ( v^{+}_{FM} )^2- (v^{+}_{CA})^2= 2 \rho_0 A v^2  k^2 $ maybe a good measure of the $ A $ term.
   However, the SF density $ \rho_s = \rho_0 v^2 $ remains  the same  as shown in Fig.1b.
   The roton $- $ mode's gap vanishes as $ \Delta_R \sim |h-h_c| $ on both sides,
   However, the roton gap at $ h > h_c $ seems independent of the $ B $ term,
   while that at $ h < h_c $ is a good measure of the $ B $ term.


{\bf 9. The nature of the quantum critical point at $ h = h_c $ and new dangerously marginally irrelevant operators }

By keeping both SF  $ ( \delta \rho _1, \theta_1) $ mode
and the critical $ \psi_2 $ mode in Eq.\ref{critical}, one can perform RG flow
on the two couplings $ \gamma_0, \gamma_1 $ between the two sectors.

     The RG analysis can be done by following the procedures developed in \cite{sfmetal} which study
     the SF to metal transition in a high temperature superconductor.
     The first line in Eq.\ref{critical} stands for the critical mode with the dynamic exponent $ z=2 $,
     one can scaling dimension of the critical field $ [ \psi_2( \vec{x}, \tau) ]= 1 $,
     then $ [u_2]=0 $ is marginally irrelevant at the upper critical dimensiona $ d_u=2 $.
     The second line stands for the SF mode with its own dynamic exponent $ z_{SF}=1 $.
     This is contrast to the case in \cite{sfmetal} where both the boson and the fermion have the same dynamic exponent $ z=2 $.
     However, to be consistent with the $ \psi_1 $ mode, one must still choose $ z=1 $.
     Then one can determine the scaling dimensions of the magnitude fluctuation $ [ \delta \rho _1( \vec{x}, \tau) ]= 2 $,
     and the phase mode $ [\theta_1( \vec{k},\omega ) ]=-3 $.
     Then the Berry phase term $  [\Gamma]=-1 $, so it becomes irrelevant, then the magnitude fluctuation $  \delta \rho _1 $
     and the phase mode are asymptotically decoupled.
     Similar phenomena happens in the two channel Kondo model \cite{kondoye123452} where the
     Berry term also becomes irrelevant and turns into a boundary condition for the Majorana fermions !
     The third line which stands for the coupling between the two sectors,
     so can determine $ [ \gamma_0]=0 $, so it is marginal.
     Similar, one can also determine $ [ \gamma_1]=0 $, so it is also marginal.

{\sl 1. The effective action }

     To construct an effective action to study the nature of the QCP,
     one can simply integrate out  $ ( \delta \rho _1, \theta_1) $ in Eq.\ref{critical}
     to study how the SF phase fluctuation affect the quantum critical behaviors of the $ \psi_2 $ mode.
     Because the $ \delta \rho _1 $ is always massive with the mass $ m_1 $, so can be integrated out to generate
     a $ |\psi_2|^4 $ term, so can be absorbed into the coefficient $ u_2 $.
     One can also see $ \langle e^{-i4 \theta_1} \rangle= \langle e^{ i4 \theta_1} \rangle=e^{-8 \langle \theta^2_1 \rangle } $
     is a finite constant at $ d=2+1 $ dimension which stands for the suppression of the BEC condensate at $ T=0 $.
     So integrating out $ ( \delta \rho_1, \theta_1 ) $ leads to the effective action describing the transition driven by  the roton dropping:
\begin{align}
    \mathcal{L}_\text{QC}
	&=\psi_2^*\partial_\tau\psi_2+v^2|\nabla\psi_2|^2 +r_2|\psi_2|^2+u_2 |\psi_2|^4    \nonumber\\
    &+\gamma_1[\psi^4_2+\psi^{*4}_2 ] + \cdots
\label{critical2}	
\end{align}
   where only the last term breaks the $ U(1) $ symmetry to the underlying $ [C_4 \times C_4]_D $ symmetry.
   However, it can be shown this symmetry breaking $ \gamma_1 $ term is marginally irrelevant
   at the $ z=2, \nu=1/2, \eta=0 $ universality class at $ d=2 $. But it is still dangerously irrelevant
   at $ h < h_c $, leads to the CAFM state in Fig.3a.

    If the Berry phase term in Eq.\ref{critical2} had became a second derivative  $ |\partial_\tau\psi_2|^2 $,
    then the dynamic exponent would be $ z=1 $,  then it is in the 3d XY universality class.
    Then the $ \gamma_1 $  maybe shown to be irrelevant near this 3d XY
    QCP by $ \epsilon= 3-d $ expansion \cite{RGflow}.
    But here it is a Gaussian fixed point with $ z=2 $, we expect its is marginally irrelevant.
    Because both $ u_2 $ and $ \gamma_1 $ are marginally irrelevant, doing RG calculations, one must treat
      $ u_2 $ and $ \gamma_1 $ on the same footing. Because the $ \gamma_1 $ term breaks the  $ U(1) $ symmetry,
      it may lead to logarithmic corrections \cite{RGflow} different from
      those just due to the $ u_2 $ term shown in Eq.\ref{tranMag}.


In Eq.\ref{critical2},at $ r_2 < 0 $,  $\psi_2$ has a condensation
$\psi_2=\sqrt{\bar{\rho}_2+\delta\rho_2}e^{i \theta_2}$,
then to quadratic order, the Lagrangian becomes
\begin{align}
    \mathcal{L}_\text{R}
	&=i\delta\rho_2\partial_\tau \theta_2
	+v^2[\frac{1}{4\bar{\rho}_2}(\nabla\delta\rho_2)^2+\bar{\rho}_2(\nabla \theta_2 )^2]\nonumber\\
	&+u_2(\delta\rho_2)^2+16\gamma_1\bar{\rho}_2^2( \theta_2)^2
\end{align}
 where we have set the linear term  $ (r_2+2u_2\bar{\rho}_2)\delta\rho_2 $ vanishing  to find $\bar{\rho}_2=-r_2/(2u_2)$.
 Then one can extract the eigen mode
$\omega_R=\sqrt{(v^2k^2+4\bar{\rho}_2u_2)(v^2k^2+16\gamma_1\bar{\rho}_2)}
\approx \sqrt{64\gamma_1\bar{\rho}_2^2u_2+\bar{\rho}_2(4u_2+16\gamma_1)v^2k^2}$
which recovers the $ \omega_{-} $ mode in Eq.\ref{lowpm}.
Note that it is the $ \gamma_1 $ term which leads to the mass of the phase $ \theta_2 $.

{\sl 2. Quantum critical scalings and quantum information scramblings in the QC regime near $ h_c $. }


The order parameter is the magnetization in the XY plane $ \langle S^x \rangle  $.
It vanishes in the Z-FM SF phase, but non-zero in the CAFM SF phase.
Because the $ \gamma_1 $ term breaks the $ U(1) $ symmetry, but is marginally irrelevant at $ d=2 $.
There is an emergent U(1) symmetry at the QCP which indicates the CAFM SF to the Z-FM SF transition
is in the same universality class of the zero-density SF-mott transition with the
critical exponents $ z=2, \nu=1/2, \eta=0 $  studied in Ref.\cite{z2}.
However, the two phases are very much different from the SF and Mott.
In the CAFM SF, there is $ \cos 4 \phi $ clock term in Eq.\ref{eq:newL} which is dangerously irrelevant near the QCP:
it is irrelevant near the QCP, but controls the quantum phase $ h < h_c $. It leads to the CAFM SF phase,
breaks the presumably $ U(1) $ symmetry to the joint $ C_4 $ symmetry and also opens a gap Eq.\ref{lowpm} to the roton mode.
There is a melting transition $ T_M $ above the CAFM SF phase ( Fig.1b ).
The universality class of melting process belongs to
2D $q=4$ state clock transition.

The standard scaling shows the roton gap should scale as $ \Delta_R \sim |h-h_c |^{z \nu} \sim |h-h_c| $
which is consistent with our specific calculations shown above.
The specific heat should scale as  $ C_R \sim T^{d/z} \sim T $.
Our specific calculations show that the specific heat due to the roton in the spin sector is
$C_\text{R}=\frac{\pi^3}{3v^2}T$.
Applying the scaling result for the $ U(1) $ conserved quantity in \cite{z2} to the order parameter
leads to :
\begin{align}
	\langle \frac{1}{N_s}\sum_i[(S_i^x)^2+(S_i^y)^2]\rangle
	=\frac{2mT}{4\pi}
	\frac{1}{\{\ln[\Lambda^2/(2mT)]\}^4}
\label{tranMag}
\end{align}
where $m=1/(2v^2)=t_0/(2t_0^2-t_\text{so}^2)$
and $\Lambda$ is a momentum upper cutoff \cite{RGflow}.

\begin{figure}[!htb]
\centering
\includegraphics[width=0.68\linewidth]{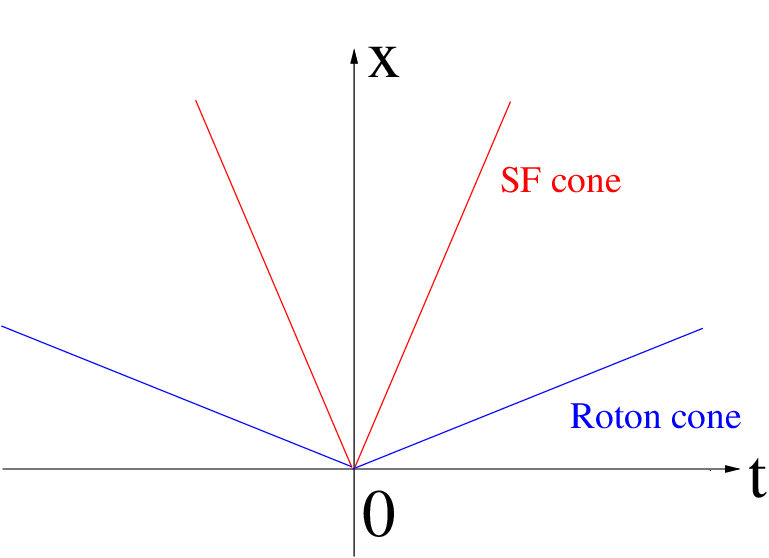}
\caption{ In the quantum information scrambling characterized by the OTOC Eq.\ref{otoc} near $ h=h_c $,
	there is a wide separation between the SF cone $ x_{sf}=v_{B,sf} t $ for $ \alpha=1 $ and the roton cone $ x_{B}=v_{B,R} t $ with $ v_{B,sf} \gg v_{B,R} $  for $ \alpha=2 $.}
\label{cone}
\end{figure}

Obviously, due to the non-linearities in Eq.\ref{blqh} and \ref{critical2}, the system shows quantum chaos \cite{SY,Kit,sun}
in the QC regime. The quantum information scrambling encoded in the spinor boson out of time ordered correlation function (OTOC):
\begin{equation}
 \langle \psi^{\dagger}_{\alpha}(t, \vec{x} ) \psi^{\dagger}_{\alpha}(0, 0) \psi_{\alpha}(t, \vec{x} )
 \psi_{\alpha}(0, 0) \rangle \sim e^{\lambda_{L,\alpha}(t- x/v_{B,\alpha})},
\label{otoc}
\end{equation}
where $ \alpha=1,2 $ stand for SF component and Roton component in Eq.\ref{boson12} respectively.
Near the lightcone $  x=v_{B,\alpha} t $ ( Fig.\ref{cone} ), the OTOC
is greatly enhanced in the QC regime in Fig.1.
The Lyapunov exponent  $ \lambda_{L,\alpha}$ due to the critical roton mode reaches its maximal value $ \lambda_{L,R} \sim T $,
while the butterfly velocity $ v_{B,R} \sim T^{1-1/z} $. For $ z=2 $,  $ v_{B,R} \sim \sqrt{T} $.
From the dimensional analysis \cite{dimen}, we conclude  $ v_{B,R} \sim  v \sqrt{T} $\cite{RGflow}.

Note that the SF sector stays un-critical across the QCP.
We obtain the finite KT transition temperature ( Fig.1b ) $T_{KT}=\frac{\pi}{2}\rho_s\sim \pi n_0v^2$.
The Lyapunov exponent due to the SF mode \cite{on} is $ \lambda_{L,sf} \sim T^3/\rho^2_s  $ and its
butterfly velocity $ v_{B,sf} \sim c= v \sqrt{ 2 n_0 U} $ as can be seen from Eq.\ref{lowpm},\ref{highpm}.

From Eq.\ref{blqh}, using Eq.\ref{spincom} and shifting the orders of the operators in Eq.\ref{otoc}, 
one may also evaluate the transverse spin-spin correlation function,
$ \langle S^{+}(t, \vec{x} ) S^{-}(0, 0)  \rangle = (-1)^{i-j}
 \langle \psi^{\dagger}_{1}(t, \vec{x} ) \psi_{2}(t, \vec{x} ) \psi^{\dagger}_{2}(0, 0) 
 \psi_{1}(0, 0) \rangle  $. Note that it can only be evaluated from the original action Eq.\ref{blqh}
 instead of the effective action Eq.\ref{critical2} where the SF model $ \psi_1 $ has been integrated out.

{\bf 10. Contrast the GL approach with  with the microscopic calculations in studying the NOFQD phenomena }

The two approaches are completely independent and complementary to each other. Both have its own
advantages and limitations.
 It is very constructive to contrast phenomenological approach in Sec.5-8  with
 the microscopic calculations in Sec.2-4.
 We can see the following mappings between the parameters in the phenomenological GL action Eq.\ref{blqh} which holds at any $ h $ and those
 evaluated by the microscopic calculations at $ h=0 $ in Sec.5 and Sec.B in the SM, especially the effective potential generated by the OFQD:
\begin{align}
 v^2&=(t_0^2-t_\text{so}^2/2)/t_0,~~r=-(\mu+4t)
      \nonumber   \\
 A&= A_m n^2,~~ B= B_m n^4,
\label{mapping}
\end{align}
  where the subscript $ m $ indicates the parameters evaluated by the microscopic calculations at $ h=0 $ in Sec.2-5.

  The interaction $ U $ and the Zeeman field $ h $ are the same in both approaches.
  Especially, it indicates that the $ A $ and $ B $ in the GL action are nothing but the effective potential generated by
  OFQD upto a density dependence. So both are positive at $ h=0 $ in the weak coupling limit.
  However, their signs may change as the interaction increases from the weak to strong couplings \cite{strongcoupling}.

  From Eq.\ref{highpm}, one can see the roton spectrum matches that achieved by microscopic calculation at its valid regime:
  high $ h $ limit. For example, the gap $ \Delta_R= 2(h-h_c)  $  matches
  the microscopic calculations in the high $ h $ limit in Sec.2. This fact may also be related to the mapping shown
  in the first row of Eq.\ref{mapping} that the GL parameter $ r $ in Eq.\ref{blqh}
  is pinned at $ r=- \rho_0 U $ when $ h \leq h_c $, but increases linearly $ r=(h-h_c)- \rho_0 U $ matches
  the chemical potential $ -(\mu+4t)=h-n_0 U $ in the microscopic calculations in Sec.2.

  The density dependence of $ A $ and $ B$  in Eq.\ref{mapping} is important in the evaluations of
  excitation spectrum at $ h >h_c $ and $ h<h_c $, especially the nature of the QCP. Had one ignored the quantum fluctuations
  in the density channel ( for example, by fixing it to be the BEC condensate $ n=n_0 $ ),
  one would have found the Zeeman field only couples to the spin sector, therefore leads to
  the un-physical physical picture: the SF sector is completely decoupled from the spin sector.
  Of course, $ A $ and $ B$ also depend on $ h $, so the mapping Eq.\ref{mapping} only holds at $ h=0 $.

  Most importantly, for the very first time, we establish the connections between the GL action and the OFQD evaluated by
  microscopic calculations !
  This observation also endows the physical meaning of the OFQD from a complete new and deep  perspective.

{\bf 11. Analogy and difference from the bilayer quantum Hall effects at the total filling factor $ \nu_T=1 $. }

It is very instructive to compare the Bilayer quantum Hall effects (BLQH) \cite{blqh1,blqh2,blqh3}
at the total filling factor $ \nu_T=1 $: Eq.\ref{blqh} is similar to the effective Chern-Simon Ginsburg-Landau (CSGL) composite boson
theory of the BLQH: in the absence of $ B $ term, the symmetry is $ U(1)_1 \times U(1)_2 = U(1)_+ \times U(1)_-$,
while the $ B $ term breaks the $ U(1)_-$ symmetry and opens a gap to the putative Goldstone mode and transfers it into
a pseudo-Goldstone mode. The $ U $ term maps to the intra-layer Coulomb iteration,
the $ A $ term maps to the inter-layer capacitive term at a finite distance $ d $.
However, there are also several  crucial differences:
(1) Here, both $ + $ and $- $ sector are charge neutral. While in BLQH,  only $ - $ is the charge neutral,
there is a Chern-Simon term in the $ + $ charge sector
which opens a charge gap to the putative SF Goldstone mode, leads to the QH with the quantized Hall conductivity $ \sigma_H=1 $
in the charge sector and the topological charges of the merons.
(2) Here, as shown in Eq.\ref{mapping} at the microscopic level,  the $ A $ term and $ B $ term are the effective potentials  generated by OFQD
in Eq.\ref{eq:Ec0cpi} which depend on the $ t/U $. However, in BLQH,
the capacitive $ A $ term  is due to the difference between intra- and inter-layers Coulomb interactions tunable by the
distance $ d $  of the two layers.
The $ B $ term is similar to the inter-layer tunneling term $ H_t=- t (\psi_1^*\psi_2 + h.c )  $ which can also
be manipulated by the in-plane magnetic field.
(3) Here the " tunneling term " between the two species $ \psi_1, \psi_2 $  is
 the 4th power dictated by the underlying $ [C_4 \times C_4]_D $  symmetry. However, it is just 1st power in BLQH.
 The power difference in the interlayer tunneling makes  dramatic differences in the two systems.
 Here, the 4th power was shown to be marginally and dangerously irrelevant near the QCP in Fig.1a, but opens the gap to the roton mode.
 However, the 1st power was known to be relevant in BLQH,, opens a gap to the interlayer SF Goldstone mode, essentially destroys the
 interlayer phase coherence.
(4) Here, there is a roton dropping transition tuned by the Zeeman field $ h $ when either $ \psi_1 $ or $ \psi_2 $
is depleted and the spin is polarized along the $ z $ direction in Fig.1a. In principle, in the absence of
the inter-layer tunneling term $ H_t $, the $ \nu_T=1 $ BLQH  is independent of the bilayer imbalance,
there is always a neutral gapless mode.
As the distance $ d $ increases, there maybe a roton dropping transition in the charge neutral sector
leading to a first order transition from the exciton superfluid (ESF) to pseudo-spin density wave (PSDW)\cite{blqh1,blqh2,blqh3}.
Both the roton mode at a finite wavevector $ |\vec{k} |=k_0 $ and SF mode near the zero wavevector $ \vec{k}=0 $ happen in the $ - $ mode,
the origin of the roton mode is due to the inter-layer interaction, so very much different from the roton mode here tuned by the
Zeeman field in Fig.3. The main difference is that the BLQH is in a a continuous system, so  the roton is at a 1d roton ring
driving a 1st order transition, while here is on a square lattice, so the roton is at $ (\pi, \pi) $  driving a 2nd order transition.
(5) Here, we did the microscopic calculations by the Bogoliubov calculations at any Zeeman field $ h $
    which can be contrasted with the phenomenological GL action Eq.\ref{blqh}. While, in the BLQH, there is a also a microscopic
    calculation called Hartree-Fock (HF) projected within the Lowest Landau level (LLL) which was also compared to
    the phenomenological Chern-Simon GL effective action \cite{blqh3}.




{\bf 12. Detectable experimental implications in cold atoms.}

%
In the ongoing experiment \cite{2dsocbec},
the BEC has $N\sim 3\times10^5$ atoms
and trapped within the diameter $d = 80\mu$m,
so one lattice site has about $n = 10$ atoms.
For typical experiment parameters,
optical lattice potential is $4.16E_r$,
and Raman potential is $1.32E_r$,
where $E_r=375$ nK denotes the recoil energy,
so the tight binding model parameter can be estimated
as $t_0\sim100$ nK, $t_\text{so}\sim 30 $ nK, so $ t_\text{so}/t_0 \sim 1/3 $.
The short-range Hubbard like interaction
$U=\frac{4\pi\hbar^2 a_s}{m}\int d^2r|w(r)|^4$,
where $s$-wave scattering length of the $^{87}$Rb atoms $a_s=103a_0$
and $a_0$ is the Bohr radius and the mass of the bosons $^{87}$Rb
lead to $U\sim 10$nK, so $ n_0 U \sim  100 $ nK.

Based on these experimental parameters,
one can estimate $T_{KT}\sim 100$ nK, the roton gap away from the critical point in Fig.3b $ \Delta_R \sim 1 $ nK,
so the meting transition $ T_M \sim 1 $ nK and the critical Zeeman field $ h_c \sim 1$ mG.
So far, the experiments are operating at $ T \sim 20 $nK.
As noted earlier,  there is also a specific heat contribution from the superfluid sector
$C_\text{sf}=\frac{\pi}{n_0Uv^2}T^2 \sim (T/n_0U) T/v^2 \ll  T/v^2 \sim C_{R} $
which is sub-leading to that from the critical roton mode in the spin sector.
Of course, the SF sector makes no contributions to the magnetization  Eq.\ref{tranMag}.
In terms of quantum information scramblings, the Lyapunov exponent due to the SF mode $ \lambda_{L,sf} \sim T^3/\rho^2_s \sim  T (T/\rho_s)^2\sim  T (T/T_{KT})^2 \ll T \sim \lambda_{L,R} $, while its butterfly velocity $ v_{B,sf} \sim c= v \sqrt{ 2 n_0 U} \gg v_{B,R}=v \sqrt{T} $.
This means the light cone of the SF mode $ x_{sf}= v_{B,sf} t $ is well within that of the roton mode $ x_{R}= v_{B,R} t $ ( Fig.\ref{cone} ).
Due to the wide separation of the two light-cones, both light-cones could be detected in the  spinor boson ( see Eq.\ref{boson12} ), spin or density OTOC.
So we conclude that the current experimental operating temperatures  may not be able to get into the CAFM SF phase,
but are still low enough such that the dramatically quantum critical fluctuation effects in
the specific heat, Lyapunov exponent, butterfly velocity,
transverse magnetization, boson or spin correlation functions can be detected
by dynamic or elastic, energy or momentum resolved,
longitudinal or transverse Bragg spectroscopies
\cite{becbragg,bragg1,bragg2},
specific heat measurements \cite{heat1,heat2}
and in-situ measurements \cite{dosexp}.

{\bf 13. Conclusions and discussions }

  In this work, we propose a new concept: NOFQD. We also perform both microscopic and
  effective GL action to implement this new concept in the specific QAH Hamiltonian Eq.1.
   We did microscopic calculations at both weak and
   strong couplings at both $ h=0 $ and $ h \gg  U $. Then we we construct the effective GL action
   at any $ h $  and contrast with the  microscopic calculations at both $ h=0 $ and very strong Zeeman field $ h \gg U $.
   However, only the GL action can capture the competition near $ h \sim h_c $.

In order to clarify the connections and relations between the conventional OFQD and the NOFQD studied here,
according to its response to a given deformation which breaks either an exact symmetry or a spurious symmetry or both,
we classify OFQD into two classes:

{\sl 1. Type-I OFQD response trivially to a deformation: }

\begin{figure}[!htb]
\includegraphics[width=0.9\linewidth]{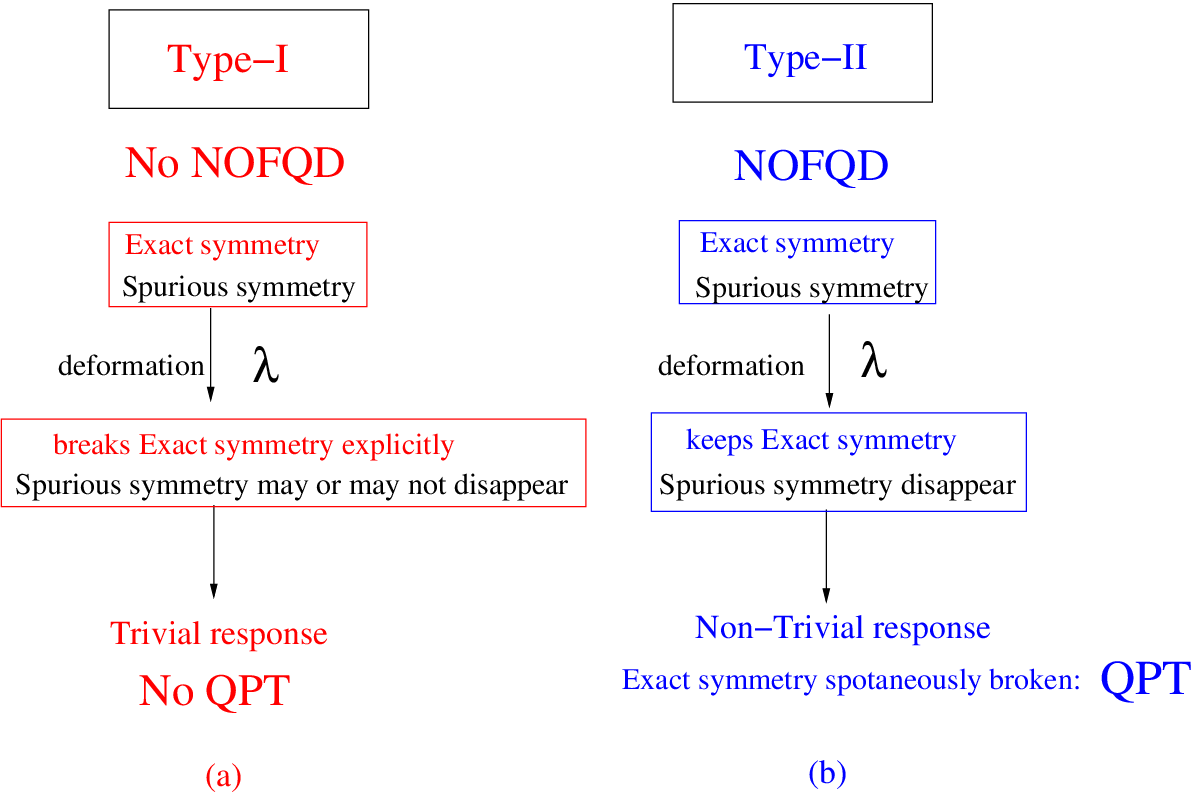}
\caption{ Contrast Type-I in (a) with type-II OFQD in (b) which response trivially and non-trivially to a deformation respectively.
 Shown are relations among the deformation, exact symmetry and spurious symmetry and their explicitly broken by the deformation $ \lambda $ or spontaneously broken through a QPT. The QPT in Type-II happens when $ \lambda $ matches the energy scale of the effective potential generated by the OFQD at $ \lambda=0 $.}
\label{ofqd}
\end{figure}

In type-I, at a high symmetry point, there is an exact symmetry and also a spurious continuous symmetry,
the OFQD breaks the spurious symmetry down to the exact symmetry and selects
the quantum ground state upto the exact symmetry.
Then a deformation $ \lambda $ away from the high symmetric point breaks the exact symmetry explicitly,
but may or may not break the spurious symmetry,
it just pick up a state from the exact degenerate manifold. If the spurious symmetry remains at $ \lambda \neq 0 $,
then the OFQD analysis at $ \lambda \neq 0 $ is very similar to that at $ \lambda=0 $ ( Fig.\ref{ofqd}a ).

{\sl 2.  Type-II OFQD response non-trivially to a deformation and leads to NOFQD }


In type-II, at a high symmetry point, there is an exact symmetry and also a spurious continuous symmetry.
The OFQD breaks the spurious symmetry down to the exact symmetry and selects
the quantum ground states upto the exact symmetry.
Then a deformation $ \lambda $ away from the high symmetric point keeps the exact symmetry, but breaks the spurious symmetry.
There is an competition between the effective potential generated by the OFQD at $ \lambda=0 $ and
the deformation $ \lambda $ as $ \lambda $ evolves,
which drives a QPT due to the spontaneously symmetry breaking of the exact symmetry.
The QPT happens when $ \lambda $ matches the scale of the effective potential.
This is what we called NOFQD: despite there is no spurious symmetry anymore when $ \lambda \neq 0 $,
but the effective potential generated at $ \lambda = 0 $ still has a dramatic effect when $ \lambda \neq 0 $  ( Fig.\ref{ofqd}b ).

It maybe constructive to compare this classification with that of superconductors in the response to an orbital magnetic field:
in Type-I, there is  a direct transition from the normal state to a superconducting state at just one critical field $ h_c $,
in Type-II, there is a mixed state intervening between the low $ h_{c1} $ and high critical field $ h_{c2} $.

{\sl 3. Two examples of Type I OFQD under SOC }

In a recent work on complete/In-complete devil stair cases in the strong coupling limit of SOC \cite{devil},
there is an exact $ [C_4 \times C_4]_D $ symmetry and also a spurious $ U(1) $ symmetry,
along the diagonal line $ \alpha=\beta $ near the Abelian point $ \alpha=\beta $ in the SOC parameter space.
The OFQD analysis selects either Y-x or X-y state as the quantum ground state which rotates to each other under
the exact $ [C_4 \times C_4]_D $ symmetry. However, slightly away from the diagonal line $ \alpha \neq \beta $, then
it  breaks the exact $ [C_4 \times C_4]_D $ symmetry explicitly,
the spurious $ U(1) $ symmetry also disappears. Then
a simple classical analysis picks up  Y-x or X-y state when $  \alpha > \beta $
or $  \alpha < \beta $ respectively. Therefore, the OFQD at $  \alpha =  \beta $ response trivially to the  deformation $  \alpha \neq  \beta $.

For another example, for the $ (\alpha=\pi/2, \beta ) $ SOC case
which has an exact $ U(1)_{soc} $ symmetry at the spin anisotropic interaction $ \lambda=1 $.
There is also a spurious $ U(1) $ symmetry at $ \lambda=1 $.
At $ \lambda=1 $,  the OFQD at the weak coupling selects the PW-X phase upto
the exact $ U(1)_{soc} $ symmetry to be the quantum ground state.
For example, the Z-x phase and the PW-X transform to each other under the exact $ U(1)_{soc} $ symmetry.
When $ \lambda < 1 $, it explicitly breaks the exact $ U(1)_{soc} $ symmetry,
but the spurious $ U(1) $ symmetry  remains,
then a similar OFQD analysis selects the PW-X phase to be the quantum ground state.
When $ \lambda > 1 $, it also explicitly breaks the exact $ U(1)_{soc} $ symmetry, the spurious $ U(1) $ symmetry also disappears,
then a simple classical analysis alone picks up the Z-x state to be the ground state.
So the OFQD at $  \lambda=1 $ response trivially to the deformation $ \lambda \neq 1 $.

Both examples belong to Type-I, the deformation breaks the exact symmetry at the high symmetric point, but it may or may not break
the spurious symmetry. There is NO competition between the effective potential generated by the OFQD
at the high symmetric point and the deformation away from the high symmetric point, no associated QPT either,
therefore no NOFQD phenomena.

{\sl 4. Two examples of Type II OFQD under SOC }

   For the QAH studied in this work,  at $ h=0 $, it has the $ {\cal M} $ symmetry and the exact $ [C_4 \times C_4]_D $ symmetry.
   At $ h=0 $ there is also a spurious $ SU(2)_s $ symmetry, an OFQD analysis selects the $ N=2 $ XY-AFM SF as the quantum ground state
   and also opens  a gap to the roton mode. This state has the $ d=4 $ degeneracy,  keeps the $ {\cal M} $ symmetry, but breaks the
   the exact $ [C_4 \times C_4]_D \rightarrow 1 $ symmetry spontaneously.
   Any small $ h \neq 0 $  breaks the $ {\cal M} $ symmetry, but still keeps the exact $ [C_4 \times C_4]_D $ symmetry.
   The spurious  $ SU(2)_s $ symmetry disappears, a simple classical analysis picks up the Z-FM SF which keeps the exact $ [C_4 \times C_4]_D $ symmetry. Then there is an associated QPT from the $ N=2 $ XY-AFM SF state to the Z-FM as $ h $ increases ( Fig.1 ).
   The exact $ [C_4 \times C_4]_D $ symmetry was spontaneously broken in the $ N=2 $ XY-AFM SF state, but restored in the Z-FM  state.
   Of course, one need to check if the 2nd QPT be pre-emptied by other competing phases.
   By both microscopic and effective GL action, we show that this does not happen .
   Despite there are many known examples of TYPE-I, we expect Type-II is also transformative to many other systems.
   For example, in a recent unpublished work \cite{haldane}, we show that Type-II also appears in the interacting bosonic Haldane model in a honeycomb lattice subject
   to a sublattice potential.

{\sl 4. Contrast to $ NAdS_2/NCFT_1 $ in SYK models }

It is very instructive to compare the NOFQD with the $ NAdS_2/NCFT_1 $ correspondence in the SYK model.
In the latter, one must move slightly away from the 1d CFT to take into accounts
the leading irrelevant operator $ -i \omega $ to study the OTOC and  extract its quantum Lyapunov exponent.
Here, a small $ h $ is also slightly away from $ h=0 $ where the OFQD selects a quantum ground state such as the $ N=2 $ XY-AFM SF state.
In the RG sense, due to the small roton gap, a small $ h $ is  also irrelevant.
However, the main difference is that in the SYK models, the ground state is a gapless QSL  state which breaks the 1d CFT ( reparamatrization ) invariance spontaneously to $ SL(2,R) $ leading to a zero ( Goldstone ) mode, the irrelevant deformation
$ -i \omega $ breaks the 1d CFT ( reparamatrization ) invariance explicitly, lifts the zero mode to a pseudo-Goldstone mode
leading to maximal chaos, but does not drive any QPT.
While here the $ N=2 $ XY-AFM SF state breaks the the exact $ [C_4 \times C_4]_D $ symmetry spontaneously and has a roton gap,
the Zeeman field $ h $ keeps the exact $ [C_4 \times C_4]_D $ symmetry, it reduces the roton gap and
just tilts the $ N=2 $ XY-AFM state into the CAFM state which has the same symmetry
as the $ N=2 $ XY-AFM SF state. However, a sufficiently large $ h $ does drive a transition to the Z-FM SF. Of course, $ NAdS_2/NCFT_1 $ correspondence is only for 1d, at higher dimension
such as $ NAdS_{d+1}/NCFT_d $ with $ d \geq 2 $, there is no need
to move slightly away from CFT, so nearly is not necessary. Here, although we demonstrate it in only $ 2+1 $ dimension, the NOFQD happens in any dimension.

As stressed in Sec.5, the original goal of constructing the GL action is to study the transition from weak coupling to strong
coupling at an integer filling. Here, we use it to investigate the transition driven by the Zeeman field $ h $
within the weak coupling regime, so the system
remains inside a SF phase. We leave our original goal to a future work \cite{strongcoupling}.
In a future work, directly inspired by the experimental realizations, we look at the deformation due to $ t_{sx} \neq t_{sy} $ which breaks
   the exact $ [C_4 \times C_4]_D $ symmetry explicitly to $ [C_2 \times C_2]_D $.  We will do  microscopic calculations at both weak and
   strong couplings.


{\bf Acknowledgements }

We thank Bao-Zong Wang and Zhan Wu for helpful discussions on experimental detections.
We also thank B. Halperin for the critical discussions in \cite{bert} and some other helpful comments on the manuscript.
We acknowledge AFOSR FA9550-16-1-0412 for supports.

\onecolumngrid
\widetext
\begin{center}
\textbf{\large Supplementary Materials for``Nearly order from quantum disorder phenomena:
its application and detection in the bosonic quantum anomalous Hall system''}
\end{center}
\setcounter{equation}{0}
\setcounter{figure}{0}
\setcounter{table}{0}
\setcounter{page}{1}
\makeatletter
\renewcommand{\theequation}{S\arabic{equation}}
\renewcommand{\thefigure}{S\arabic{figure}}
\renewcommand{\bibnumfmt}[1]{[S#1]}
\renewcommand{\citenumfont}[1]{S#1}

In the main text, we present an NOFQD analysis on
the weakly coupling bosonic Quantum Anomalous Hall model leading to the phase diagram in Fig.M1b.
In this Supplementary Materials (SM),
in the first section, we present a traditional approach which ignores the NOFQD phenomena  and leads to a first order transition at $ h=0 $.
The breakdown of this traditional approach at $ h \lesssim U $ was used to motivate the importance of the NOFQD phenomena
and also inspire us to develop the
new symmetry based GL approach in the main text to study the  NOFQD phenomena in the most systematic and complete way.
In section two, we provide the explicit form of the
effective potential Eq.M15  generated by the OFQD analysis at $ h=0 $
 which was also predicted by the independent symmetry based GL approach.

\section{ A. The conventional approach which ignores the NOFQD phenomena leads to a first order transition at $ h=0 $.  }

In the weak coupling limit, the conventional approach is to treat the kinetic ( single particle ) term exactly, then
treat the weak interaction term by perturbation theory. In the opposite strong coupling limit, one may treat interaction exactly, then
treat the kinetic term by perturbation theory \cite{rh}. However, in the intermediate regime, usually no controlled analytical approach can be
found.

In this manuscript, we are focusing on the weak interaction $ U < t_0 $ regime, which
is also the experimentally relevant situation. So conventional approach can be used to study  the Z-FM in the $ h \gg U $ limit.
However, when $ h $ becomes also weak, namely,
in the regime $ h \lesssim U $, it becomes quite tricky and difficult to study the delicate balance between the
weak interaction effects and the weak Zeeman effects.
Here, we show that if one still takes the conventional ( or traditional ) approach which treats the kinetic ( single particle ) term first,
then the interaction by perturbation, one reaches that there is a first order transition line
at $ h=0 $ ( Fig.S1a ). As shown in the main text, this conclusion is in-correct. This is because
the conventional approach fails to capture the NOFQD effect which
treats the kinetic term and interaction effects on the equal footing, so
completely breaks down in this regime. A new and systematic approach  is needed to capture the NOFQD effect.

Such a new approach is the GL action  developed in the main text.
As alerted in the first paragraph, in a general many body interacting system, it is very difficult to construct
a controlled analytic approach  when the two competing scales such as the kinetic energy and the interaction are comparable.
Here, we make two crucial observations (1) even in the presence of $ h $, the quantum ground state can still be chosen from the manifold spanned by the family of states shown in Eq.M2 parameterized by Eq.M4.
(2) We managed to find an analytic formula for the effective potential generated by the order from disorder phenomena Eq.M5
which holds in the whole range $ ( 0< \theta < \pi, 0 < \phi < 2 \pi) $.
 ( For the explicit derivation of this formula, see the next section of the SM ).
 This complete formula is important, because the minimum $ (\theta_0, \phi_0 ) $ changes as $ h $ increases.
 In fact, this complete formula was also predicted  independently from the symmetry based GL action.

It is these two facts which enable us to democratically and systematically
capture the NOFQD effect which stands for the
delicate competition between the effective potential generated by the OFQD due to the weak interaction
and the Zeeman energy due to a small finite $ h $, therefore map out the global phase diagram Fig.M1.
Especially, the first order transition at $ h=0 $ is transferred to the
two second order transitions at $ h= \pm h_c $,
then a "new" phase called canted AFM phase ( CAFM ) emerges in the narrow window of $ 0 < h < h_c $
( it still has the same symmetry as the $ N=2 $ XY-AFM phase ).
Obviously, there is a qualitative change from first order transition reached by the conventional approach
whcih ignores the NOFQD effects and the
second order transition reached by the new approach which incorporates the NOFQD effects.
It is the second order transition associated with the  NOFQD  which leads
to all the sharp experimentally detectable  predictions presented in the main text.

Although this conventional approach breaks down and lead to in-correct physics at $ h \lesssim U $, especially  a first order transition at $ h=0 $,
it is still instructive to present this traditional approach in this SM, because
it was used to inspire the authors to discover the NOFQD phenomena and to develop the new and systematic GL action approach in the main text to explore the NOFQD phenomena \cite{bert}.
It may also help to  shed some lights on the physical implications of NOFQD phenomena on the phase diagram in Fig.M1.

{\sl (a) The conventional approach to find the Z-FM at any $ h >0 $. }

Any small $ h \neq 0 $ breaks the mirror symmetry and
split the degenerate single-particle state
at momentum $(0,0)$ and $(\pi,\pi)$.
A straightforward mean field analysis predicts the condensation occurs at
either $(0,0)$ or $(\pi,\pi)$ depending on the sign of $h$.
the unique lowest energy single-particle state lead to
$|\Psi_{\text{sf},\uparrow}\rangle
\sim(\sum_i a_{i\uparrow}^\dagger)^N|0\rangle$ if $h>0$
or $|\Psi_{\text{sf},\downarrow}\rangle
\sim[\sum_i (-1)^{i_x+i_y}a_{i\downarrow}^\dagger]^N|0\rangle$ if $h<0$.
Due to their spin alinement along the $ z $ axis, we name it $ Z-FM  $ superfluid.

Then after choosing  $(0,0)$ as the unique BEC condensation point at $ h >0 $, we
apply  the Bogoliubov theory to study the effects of the interaction $ U $ at any $ h \neq 0 $.
As shown in Sec.M2,
we obtain a linear gapless superfluid mode and a gapped roton mode with the gap at $ \vec{Q}= (\pi,\pi) $ listed in
Eq.\ref{rotonh}
\begin{equation}
    \Delta^{0}_R(h)=2|h|     \nonumber
\end{equation}
 shown in Fig.S1b (  also shown in the dash line in Fig. M4b in the main text ).
 As one decreases $h$ from $ h \gg U $ to $ h \lesssim U $, then to $0^{+}$,
the roton mode at $ \vec{Q}= (\pi,\pi) $ gets lower and lower, then touches zero at $ h=0^{+} $.
This behaviour may signify a possible first order transition at $h_c=0^{+}$.
This physical picture leads to the naive phase diagram shown in Fig.S1a.

As shown in the main text, this conventional approach is valid at $ h > U $, but breaks down when $ h \lesssim U $.
Then the next two steps

{\sl (b) The order from quantum disorder analysis to find the $ N=2 $ XY-AFM state at $ h=0 $  }
and

{\sl (c) The Roton mass gap generated from the "order by disorder" mechanism in the $ N=2 $ XY-AFM state st $ h=0 $. }
were performed in Sec.3 and Sec.4 in the main text respectively.

{\sl (d) A naive ( also wrong ) Roton mass gap at $ h> 0 $ which ignores the NOFQD effects. }

  In the conventional approach taken in (a), we treat $ h $ exactly, then treat $ U $ by perturbation. So
  we may expect that the roton gap at any $ h \neq 0 $ could be obtained simply by
  incorporating the roton gap Eq.\ref{roton0} at $ h=0 $  directly into the roton gap Eq.\ref{rotonh} at $ h \neq 0 $:
\begin{equation}
 \Delta_R(h)=\sqrt{\Delta^2_R( h=0 )+ 4h^2 }
\label{rotonh0}
\end{equation}
  which is always above the dashed line and shown in the solid line in Fig.S1b.

\begin{figure}[!htb]
\includegraphics[width=15cm]{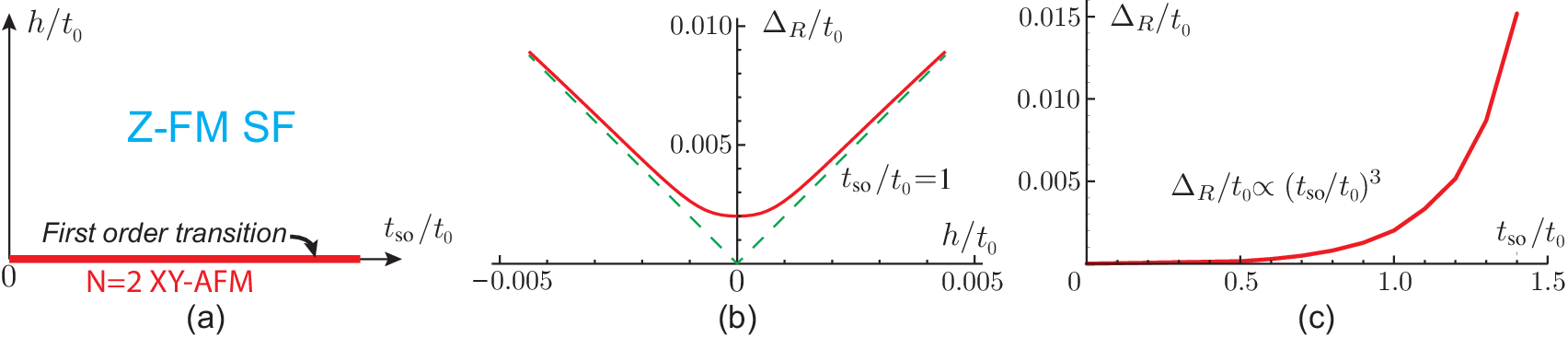}
\caption{(a) The naive zero temperature phase diagram as a function of
	$t_\text{so}/t_0$ and $h/t_0$ with fixed $n_0U$ reached by the conventional approach which ignores the NOFQD effects.
    If ignoring the  NOFQD effects, then any small $ h > 0 $ will select the Z-FM.
    So any small $ h $ would induce a first order transition from the $ N=2 $ XY-AFM to the Z-FM.
    However, incorporating the NOFQD effects
    will change the naive phase diagram to the correct Fig.1a in the main text.
    (b) The dashed line indicates that  when the OFQD effect at $ h=0 $  is ignored,
    the roton gap drops to zero as $ h \to 0^{+} $ as Eq.\ref{rotonh}.
    Then the OFQD  generates the roton gap Eq.\ref{roton0} at $ h=0 $ whose dependence on $ t_\text{so}/t_0 $  is given in (c).
    The solid line is the naively extrapolated roton gap behaviours Eq.\ref{rotonh0} in the conventional approach.
    (c)  At $ h=0 $, the Roton gap computed by the OFQD analysis
        is an increasing function of  $t_\text{so}$ as approaching to the right.
	Other parameters are $n_0U=1$, $n_0\approx1$.
        At $ t_\text{so}=0 $, it is nothing but the FM spin wave mode.
        At small $ t_\text{so} $, it can also be calculated by the perturbation theory in $ t_\text{so} $
        which leads to $ \Delta_R \sim t^{3}_\text{so} $ ( see the next section below Eq.\ref{eq:pb} ).
        As shown in the main text, only (c) remains correct, (a) and (b) should be replaced by Fig.M1a and Fig.M4b respectively.  }
\label{QCP}
\end{figure}

  Eq.\ref{rotonh0} leads to the following scenario:
  The transition at $ h=0 $ is a first order phase transition in Fig.S1a driven by the roton dropping tuned by the Zeeman field $ h $ shown in the solid line in Fig.S1b.
  The roton still has a non-zero gap $ \Delta_R( h=0 ) $ in Eq.\ref{roton0} before it sparks the first order transition.

 {\sl  Now the contradiction comes: if it were indeed a first order transition at $ h=0 $, then the state would be a mixed state
  of the $ (0,0) $ spin up state and $ (\pi,\pi) $ spin down state
  with any ratio. However, by the OFQD phenomena at $ h=0 $, we found the state is a pure state: the $ N=2 $ XY-AFM state
  instead of a mixed state. This contradiction shows the conventional approach may break down when $ h \to 0^{+} $. }


Indeed, as shown in the main text, this naive scenario holds only when $ h $ is sufficiently large $ h > U $,
but breaks down at a small $ h \lesssim U $. As explicitly demonstrated in the main text, to compute the roton spectrum at
$ h \lesssim U $, one must examine the NOFQD effects which
treat the effects of $ U $ and $ h $ on equally footing.
Namely, by treating the two nearly degenerate states at $ (0,0 ) $ and $ (\pi,\pi) $ on equal footing, one can
investigate the competitions between the effective potential  generated by the OFQD in Fig.M2 due to the
interaction $ U $ and the Zeeman energy due to a small finite $ h $.
The systematic GL effective calculation developed in the main text precisely capture the NOFQD effects and
leads to the correct roton spectrum
in the second equation in Eq.M18 where one can extract the correct roton gap at any $ h< h_c $:
\begin{equation}
\Delta^{-}_R=4AB\sin^4\theta_0 =2\sqrt{B/A^3}(A^2-n_0^2h^2)/n_0
\label{rotoncorrect}
\end{equation}
 which is listed in Eq.\ref{RR} and shown in the solid line in Fig.M4b.
 In the $h\to h^{-}_c$ limit, the roton gap vanishes as $\Delta_R\sim h_c-h$.
 Note that the new  and systematic GL action approach not only leads to the correct roton gap Eq.\ref{rotoncorrect}, but also
 the correct roton spectrum at any $ k $ in the long wavelength limit in
 the second equation in Eq.M18 ( the solid line in Fig.M3b) which is qualitatively different from
 the naively speculated roton spectrum Eq.\ref{rotonh0} ( the solid line in Fig.S1b ).

Therefore, the breakdown of the conventional approach at $ h \lesssim U $ leads to a qualitative change of the physical picture:
The first order transition at $ h=0$ splits into two second order transitions at
a finite $ h = \pm h_c $ shown in Fig.M1a. The $ N=2 $ XY-AFM SF phase at $ h=0 $ turns into the canted AFM SF phase at $ h >0 $.
Of course, the two phases still have the same symmetry breaking patterns. It is the zero temperature 2nd order transition
associated with the NOFQD effects which leads to all the sharp predictions on
all the experimental measurable quantities analyzed in the main text.

In short, we demonstrate that the effective potential generated from the OFQD
at least has 3 important impacts in the bosonic QAH system:
(a) at $ h=0 $, it leads to a quantum ground state selection rule,
(b) it also opens a gap to the roton mode at $ h=0 $
(c) Most importantly, at small $ h \lesssim U $, it leads to the NOFQD effects.
Then  we develop a new  and systematic GL effective action scheme to capture the NOFQD effects which stands for the competition
between the effective potential generated by the "order from disorder" mechanism with the Zeeman energy due to a small finite $ h $.
The competition leads to the corrected roton excitation spectra at any $ k $ in the long wavelength limit in the second equation in Eq.M18.
It is the roton gap closing at $ h=h_c $ which
drives a second order quantum phase transition (QPT) from the CAFM to the Z-FM in Fig.M1b.
The QPT at $ T=0 $ associated with the NOFQD effects leads to sharp and unique predictions  in several
physical quantities which can be easily detected in the current experiments.




\section{ B. The functional form of the effective potential generated by the OFQD }

  Here, we will explicitly demonstrate the explicit functional form Eq.M5 which holds in the whole range $ 0< \theta < \pi, 0 < \phi < 2 \pi) $.
  This complete formula is important, because the minimum $ (\theta_0, \phi_0 ) $ changes as $ h $ increases.
It is this formula which enable us to democratically and systematically
capture the NOFQD effects, therefore map out the global phase diagram Fig.M1.
Most importantly,  this formula reached by the microscopic calculation at weak coupling precisely match
the most general form in the  phenomenological GL action
constructed just from the symmetry principle. This match establishes the deep connection between
the two independent, different, but complementary approaches.

\subsection{1. Analytical approach: perturbation theory in $ t_\text{so}/t_0 $. }

In Fig.M1a, at small $ t_\text{so}/t_0 $ and $ h=0 $, we perform the perturbation calculation
in $ t_\text{so}/t_0 $ which is independent and complementary to the order from disorder analysis done in the main text.
This in-dependent approach can also be used as a check on the results achieved from the order from disorder analysis.

Before performing perturbation calculations,
it is convenient to apply transformation
$a_{i\downarrow}\to (-1)^{i}a_{i\downarrow}$
to the original Hamiltonian Eq.(1)
and obtain
\begin{align}
    \mathcal{H}
	=-t_0\sum_{\langle ij\rangle}a_i^\dagger a_j
	+it_\text{so}\sum_{\langle ij\rangle}(-1)^i
	a_i^\dagger(d_{ij}\cdot\vec{\sigma})a_j
	-h\sum_i a_i^\dagger\sigma^z a_i
	+\frac{U}{2}\sum_i n_i(n_i-1)
	-\mu\sum_in_i
\end{align}
Since the ground-state energy is a gauge invariant quantity,
we can use the Hamiltonian Eq.(S1) for the ground-state energy calculation.
In the absent of Zeeman field $ h=0 $,
we first separate Hamiltonian Eq.(S1) into two parts
\begin{align}
    \mathcal{H}_0
	=-t_0\sum_{\langle ij\rangle}a_i^\dagger a_j
	+\frac{U}{2}\sum_i n_i(n_i-1)
	-\mu\sum_in_i,\quad
    \mathcal{H}_\text{so}
	=it_\text{so}\sum_{\langle ij\rangle}
	(-1)^ia_i^\dagger(d_{ij}\cdot\vec{\sigma})a_j
\end{align}

Becasue the Hamiltonian $\mathcal{H}_0$ has the spin $ SU(2)_s $ symmetry in the rotated $ \tilde{SU}(2) $ basis
$\tilde{\mathbf{S}}_i=((-1)^{i_x+i_y}S_i^x,(-1)^{i_x+i_y}S_i^y,S_i^z)$,  at the weak coupling $U\ll t$
one can condense the spinor bosons at $ \bf{k}=0 $ with the spin symmetry breaking direction along $ (\theta, \phi) $.
So the ground-state is a superfluid with many-body wavefunction
$|\Psi_\text{sf}\rangle\sim(\sum\Psi_ia_i^\dagger)^N|\text{Vac}\rangle$
where $\Psi_i=c_0 \eta_0+c_\pi \eta_\pi$
and $|c_0|^2+|c_\pi|^2=1$.
Notice we are useing the same notation $\eta_0=(\begin{smallmatrix}1\\0\end{smallmatrix})$ and
$\eta_\pi=(\begin{smallmatrix}0\\1\end{smallmatrix})$ as defined in main text.
The coefficients  parameterize the same  as Eq. M4  $c_0=e^{-i\phi/2}\cos(\theta/2)$
and $c_\pi=e^{+i\phi/2}\sin(\theta/2)$.

Within Bogoliubov approximation,
the original boson operators is replaced by condensate part plus a fluctuations
$a_i=\sqrt{N_0}\Psi_i+\psi_i$,
thus one can expand $\mathcal{H}_0$
and keep terms up to second order in the fluctuations.
The quadratic theory of $\mathcal{H}_0$ after the  expansion is $\mathcal{H}_{0,\text{Bog}}$,
which can be diagonalized as
\begin{align}
    \mathcal{H}_{0,\text{Bog}}
	=const.+\sum_\mathbf{k}(
	    \omega_{1,\mathbf{k}}\alpha_{1,\mathbf{k}}^\dagger\alpha_{1,\mathbf{k}}
	    +\omega_{2,\mathbf{k}}\alpha_{2,\mathbf{k}}^\dagger\alpha_{2,\mathbf{k}})
\end{align}
where Bogoliubov transformation is given by
\begin{align}
    \psi_{\mathbf{k}\uparrow}
	=-\bar{c}_\pi\alpha_{1,\mathbf{k}}
	    +c_0(\cosh\xi_\mathbf{k} \alpha_{2,\mathbf{k}}
		+\sinh\xi_\mathbf{k} \alpha_{2,-\mathbf{k}}^\dagger),\quad
    \psi_{\mathbf{k}\downarrow}
	=+\bar{c}_0\alpha_{1,\mathbf{k}}
	    +c_\pi(\cosh\xi_\mathbf{k} \alpha_{2,\mathbf{k}}
		+\sinh\xi_\mathbf{k} \alpha_{2,-\mathbf{k}}^\dagger)
\end{align}
and the excitation spectrum are $\omega_{1,\mathbf{k}}=4t_0-2t_0(\cos k_x+\cos k_y)$
and $\omega_{2,\mathbf{k}}=\sqrt{\omega_{1,\mathbf{k}}(\omega_{1,\mathbf{k}}+2n_0U)}$.
The auxiliary angle is defined as $\xi_\mathbf{k} =-\frac{1}{2}\arctanh[\frac{\omega_{1,\mathbf{k}}}{\omega_{1,\mathbf{k}}+n_0U}]$.
The ground-state $|0\rangle$ of $\mathcal{H}_{0,\text{Bog}}$ is  the vacuum of $\alpha_{1,\mathbf{k}}$ and $\alpha_{2,\mathbf{k}}$,
so that $\alpha_{1,\mathbf{k}}|0\rangle=0$ and $\alpha_{2,\mathbf{k}}|0\rangle=0$.

The $ SU(2)_s $ symmetry of $\mathcal{H}_{0}$ guaranteed that $c_0$ and $c_\pi$ can take any values
as long as they satisfy normalization condition.  Namely, any spin symmetry breaking direction along $ (\theta, \phi) $ is equivalent.
However, $\mathcal{H}_\text{so}$ explicitly breaks the $ SU(2)_s $ symmetry,
so the $ SU(2)_s $ symmetry becomes spurious and only exists at mean-field level.

Now the perturbation $\mathcal{H}_\text{so}$ takes following form in the Bogoliubov approximation
\begin{align}
    \mathcal{H}_\text{so}
	=2t_\text{so}\sum_\mathbf{k}
	(\gamma_\mathbf{k}
	\psi_{\mathbf{k}\uparrow}^\dagger
	\psi_{\mathbf{k}+\mathbf{Q}\downarrow}+h.c.)
\end{align}
where $\gamma_\mathbf{k}=\sin k_x-i\sin k_y$
and $\mathbf{Q}=(\pi,\pi)$.

In order to apply the perturbation,
it is convenient to express $\mathcal{H}_\text{so}$
in terms of $\alpha_{\mathbf{k}}$:
\begin{align}
    \mathcal{H}_\text{so}
	&=t_\text{so}\sum_\mathbf{k}
	\big
	[f_{1,\mathbf{k}} \alpha_{1,\mathbf{k}}^\dagger\alpha_{1,\mathbf{k+Q}}
	+f_{2,\mathbf{k}} \alpha_{2,\mathbf{k}}^\dagger\alpha_{2,\mathbf{k+Q}}
	+f_{3,\mathbf{k}} \alpha_{1,\mathbf{k}}^\dagger\alpha_{2,\mathbf{k+Q}}
	+f_{4,\mathbf{k}} \alpha_{2,\mathbf{k}}^\dagger\alpha_{1,\mathbf{k+Q}}    \nonumber\\
	&+(f_{5,\mathbf{k}} \alpha_{1,\mathbf{k}}^\dagger\alpha_{2,\mathbf{-k+Q}}^\dagger
	 +f_{6,\mathbf{k}} \alpha_{2,\mathbf{k}}^\dagger\alpha_{2,\mathbf{-k+Q}}^\dagger
	 +h.c.)\big]
\label{hso}
\end{align}
where the coefficients are
\begin{align}
    f_{1,\mathbf{k}}
	&=-\sin\theta(e^{i\phi}\gamma_\mathbf{k}-e^{-i\phi}\bar{\gamma}_\mathbf{k})\\
    f_{2,\mathbf{k}}
	&=\sin\theta(\cosh\xi_\mathbf{k}\cosh\xi_\mathbf{k+Q}+\sinh\xi_\mathbf{k}\sinh\xi_\mathbf{k+Q})
	    (e^{i\phi}\gamma_\mathbf{k}-e^{-i\phi}\bar{\gamma}_\mathbf{k})\\
    f_{3,\mathbf{k}}
	&=-2\cosh\xi_\mathbf{k+Q}
	    [e^{i\phi}\sin^2(\theta/2)\gamma_\mathbf{k}
	    +e^{-i\phi}\cos^2(\theta/2)\bar{\gamma}_\mathbf{k}]\\
    f_{4,\mathbf{k}}
	&=2\cosh\xi_\mathbf{k}
	    [e^{i\phi}\cos^2(\theta/2)\gamma_\mathbf{k}
	    +e^{-i\phi}\sin^2(\theta/2)\bar{\gamma}_\mathbf{k}]\\
    f_{5,\mathbf{k}}
	&=-2\sinh\xi_\mathbf{k+Q}
	    [e^{i\phi}\sin^2(\theta/2)\gamma_\mathbf{k}
	    +e^{-i\phi}\cos^2(\theta/2)\bar{\gamma}_\mathbf{k}]\\
    f_{6,\mathbf{k}}
	&=\sin\theta \cosh\xi_\mathbf{k}\sinh\xi_\mathbf{k+Q}
	(e^{i\phi}\gamma_\mathbf{k}-e^{-i\phi}\bar{\gamma}_\mathbf{k})
\end{align}

By treating $t_\text{so}/t_0$ as a small parameter,
one can apply perturbation theory to obtain
\begin{align}
    E_\text{ofd}=\delta E^{(1)}+\delta E^{(2)}
	+\delta E^{(3)}+\delta E^{(4)}+\mathcal{O}(t_\text{so}^5)
\end{align}
where we denote $\delta E^{(n)}\propto t_\text{so}^n$.
Since $\mathcal{H}_\text{so}$ has odd parity,
so all odd order terms vanish and
\begin{align}
    \delta E^{(2)}&=
	\langle 0|\mathcal{H}_\text{so} g \mathcal{H}_\text{so}|0\rangle\\
    \delta E^{(4)}&=
	\langle 0|\mathcal{H}_\text{so}g
		  \mathcal{H}_\text{so}g
		  \mathcal{H}_\text{so}g
		  \mathcal{H}_\text{so}|0\rangle
	-\langle0|\mathcal{H}_\text{so}g
		  \mathcal{H}_\text{so}|0\rangle
	 \langle0|\mathcal{H}_\text{so}g^2
		  \mathcal{H}_\text{so}|0\rangle
\end{align}
where $g=\sum_{n\neq0}\frac{|n\rangle\langle n|}{-E_n}$.

After a substitution of  $\mathcal{H}_\text{so}$ in Eq.\ref{hso}, we reach
\begin{align}
    \delta E^{(2)}
	&=-t_\text{so}^2\sum_k\left[
		\frac{|f_{5,\mathbf{k}}|^2}{\omega_{1,\mathbf{k}}+\omega_{2,\mathbf{Q-k}}}
		+\frac{|f_{6,\mathbf{k}}|^2+f_{6,\mathbf{k}}\bar{f}_{6,\mathbf{Q-k}}}
		      {\omega_{2,\mathbf{k}}+\omega_{2,\mathbf{Q-k}}}
	\right]
\end{align}
and
\begin{align}
    \delta E^{(4)}
	=-t_\text{so}^4\sum_\mathbf{k}&\left[
	\frac{|f_{3,-\mathbf{k}}f_{5,\mathbf{k}}|^2}
	     {\omega_{1,\mathbf{k}}(\omega_{1,\mathbf{k}}+\omega_{2,\mathbf{Q+k}})^2}
	+\frac{1}{\omega_{2,\mathbf{k}}}\left|
	\frac{f_{4,\mathbf{Q+k}}f_{5,\mathbf{Q+k}}}
	     {\omega_{1,\mathbf{Q+k}}+\omega_{2,\mathbf{k}}}
	+
	\frac{f_{2,-\mathbf{k}}(f_{6,\mathbf{k}}+f_{6,\mathbf{Q-k}})}
	     {\omega_{2,\mathbf{k}}+\omega_{2,\mathbf{Q-k}}}
	\right|^2
	\right. \nonumber\\
	&\left.
	+
	\frac{1}{\omega_{1,\mathbf{k}}+\omega_{2,\mathbf{k}}}
	\left|
	\frac{f_{1,\mathbf{k}}f_{5,\mathbf{Q+k}}}{\omega_{1,\mathbf{Q+k}}+\omega_{2,\mathbf{k}}}
	+
	\frac{f_{2,-\mathbf{k}}f_{5,\mathbf{Q+k}}}{\omega_{1,\mathbf{k}}+\omega_{2,\mathbf{Q-k}}}
	+
	\frac{f_{3,\mathbf{k}}f_{6,-\mathbf{k}}}{\omega_{2,-\mathbf{k}}+\omega_{2,\mathbf{Q+k}}}
	+
	\frac{f_{3,\mathbf{k}}f_{6,\mathbf{Q+k}}}{\omega_{2,\mathbf{Q+k}}+\omega_{2,\mathbf{k}}}
	\right|^2
	\right]
	+\cdots
\end{align}
where $\cdots$ means $\phi$-independent part.

After some manipulations, we found
\begin{align}
	\delta E^{(2)}&=t_\text{so}^2N_s[c_2+a(1+\cos2\theta)],\\
	\delta E^{(4)}&=t_\text{so}^4N_s[c_4+b\sin^4\theta(1+\cos4\phi)+\cdots],
\label{e2e4}
\end{align}
where $\cdots$ also contains $\theta$-only dependent terms like $\cos2\theta$ and $\cos4\theta$ which is subleading to $\delta E^{(2)}$ when $t_\text{so}\ll t$.
The coefficients are
\begin{align}
	a&=\frac{1}{N_s}\sum_\mathbf{k}|\gamma_\mathbf{k}|^2\Bigg(
	\frac{\cosh^2\xi_\mathbf{k}\sinh^2\xi_\mathbf{Q+k}
	+\cosh\xi_\mathbf{k}\sinh\xi_\mathbf{k}\cosh\xi_\mathbf{Q+k}\sinh\xi_\mathbf{Q+k}}
	{\omega_{2,\mathbf{k}}+\omega_{2,\mathbf{Q+k}}}
	-\frac{\sinh^2\xi_\mathbf{Q+k}}
	     {\omega_{1,\mathbf{k}}+\omega_{2,\mathbf{Q+k}}}\Bigg)\\
	b&=\frac{1}{N_s}\sum_\mathbf{k}(\gamma_{\mathbf{k}}^4+\bar{\gamma}_{\mathbf{k}}^4)\Bigg[
	-\frac{\sinh^2(2\xi_{\mathbf{k}+\mathbf{Q}})}{4\omega_{1,\mathbf{k}}}
	-\frac{1}{4\omega_{2,\mathbf{k}}}
	\Big(\frac{\sinh(2\xi_\mathbf{k})}
		  {\omega_{1,\mathbf{k}+\mathbf{Q}}+\omega_{2,\mathbf{k}}}
	-\frac{\sinh(2\xi_\mathbf{k}+2\xi_{\mathbf{k}+\mathbf{Q}})}
	      {\omega_{2,\mathbf{k}+\mathbf{Q}}+\omega_{2,\mathbf{k}}}\Big)^2 \nonumber\\
	&\qquad\quad
	+\frac{1}{\omega_{1,\mathbf{k}}+\omega_{2,\mathbf{k}}}
	\Big(\frac{\sinh(\xi_{\mathbf{k}})}
	    {\omega_{1,\mathbf{k}+\mathbf{Q}}+\omega_{2,\mathbf{k}}}
	+\frac{\sinh(\xi_{\mathbf{k}+\mathbf{Q}})
		\cosh(\xi_\mathbf{k}+\xi_{\mathbf{k}+\mathbf{Q}})}
	      {\omega_{1,\mathbf{k}}+\omega_{2,\mathbf{k}+\mathbf{Q}}}
	+\frac{\cosh(\xi_{\mathbf{k}+\mathbf{Q}})
		\sinh(\xi_\mathbf{k}+\xi_{\mathbf{k}+\mathbf{Q}})}
	      {\omega_{2,\mathbf{k}}+\omega_{2,\mathbf{k}+\mathbf{Q}}}
	\Big)^2\Bigg]
\end{align}

For $U>0$ case, we  can prove that $a,b>0$.
So $\delta E^{(2)}$ and $\delta E^{(4)}$
reach the minima at $\cos2\theta=-1$ and $\cos4\phi=-1$ respectively,
which tells the quantum ground-state sits at $\theta=\pi/2$ and $\phi=\pi/4$ which are nothing but the XY-FM state.
Transforming it back from the rotated $ \tilde{SU}(2) $ basis to the original basis, it is nothing but the $ N=2 $ XY-AFM SF state.
So we reach the same ground state from the
two independent methods.

 When comparing Eq.\ref{e2e4} with Eq.M5,
\begin{align}
    E_\text{ofd}
	=E_{\text{ofd},0}
	+N_s\Big[\frac{A}{4}(1+\cos2\theta)
	+\frac{B}{16}\sin^4\theta(1+\cos4\phi)\Big]\> + \cdots
\label{eq:pb}
\end{align}
   where the $ \cdots $ means the higher order terms respecting the same symmetry of the Hamiltonian.

Then we obtain the analytical expressions of $A= a (2t_\text{so})^2$
and $B= b (2t_\text{so})^4$ from the perturbation theory.
Then Eq.\ref{roton0} leads to $ \Delta_R \sim t^{3}_s $ at $ h=0 $ shown in Fig.S1b.
As shown in the main text, Eq.\ref{eq:pb} acn be written as the form in the GL action:
\begin{align}
\frac{A}{2}(|\psi_1|^2-|\psi_2|^2)^2+\frac{B}{2}[(\psi_1^*\psi_2)^2+(\psi_1\psi_2^*)^2]^2 +
C(|\psi_1|^2-|\psi_2|^2)^4 + D [(\psi_1^*\psi_2)^2+(\psi_1\psi_2^*)^2]^4  + \cdots
\label{eq:pb1}
\end{align}
  where $ \cdots $ means the even higher order terms than $ C $ and $ D $ term
  respecting the $ [ C_4 \times C_4 ]_D $ symmetry of the Hamiltonian. In fact, as shown below, the $ C $ and $ D $ terms can also be
  determined numerically by the microscopic calculations. Both found to be positive.

\subsection{2. Numerical approach: compared with the analytical results}

We will show that although the form Eq.\eqref{eq:pb}
is obtained from the perturbation theory in a small $ t_\text{so}/t_0 $,
its form fits numerical data very well even upto $ t_\text{so}/t_0=1$ .
The data is obtained from numerical evaluation
$\omega_{l,\mathbf{k}}$ and
$E_\text{ofd}(\theta,\phi)$ in Eq.M3.

The comparison consist of two parts:
the global property and the local property.

For the global property,
we plot numerical results from Eq.M3
and the analytical result from Eq.\eqref{eq:pb} in Fig.S2.
It is clear that the two results fit quite well in the whole range $ 0< \theta < \phi, 0 < \phi< \pi/2 $.

On the other hand, if we only keep the form of the perturbation theory
and treat coefficients $A$ and $B$ as fitting parameters,
then the relative error can be controlled below 1\%.
Figure S2 shows the form work very well
even for $t_\text{so}$ is not small,
i.e. $t_\text{so}/t=1$.

\begin{figure}[!htb]
\centering
    \includegraphics[width=\linewidth]{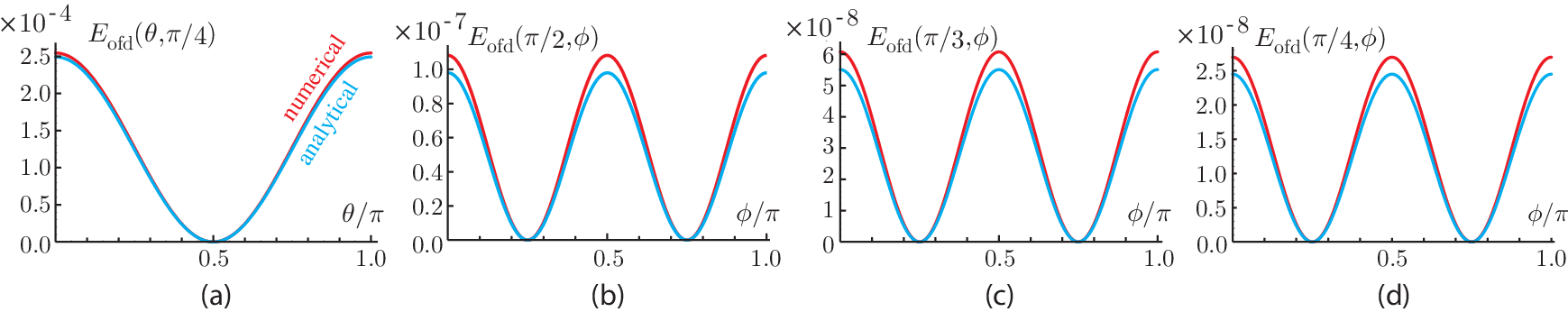}
    \caption{Plot of $E_\text{ofd}$ for varies parameter regimes
	(a) $0<\theta<\pi$ and $\phi=\pi/4$,
	(b) $\theta=\pi/2$ and $0<\phi<\pi$,
	(c) $\theta=\pi/3$ and $0<\phi<\pi$,
	(d) $\theta=\pi/4$ and $0<\phi<\pi$.
	The numerical results are represented by the red lines,
	and the analytical results from the perturbation theory
	are represented by the blue lines.
	The other parameters are $t_\text{so}/t=1/3$, $n_0U/t_0=1$.}
\end{figure}

\begin{figure}[!htb]
\centering
    \includegraphics[width=\linewidth]{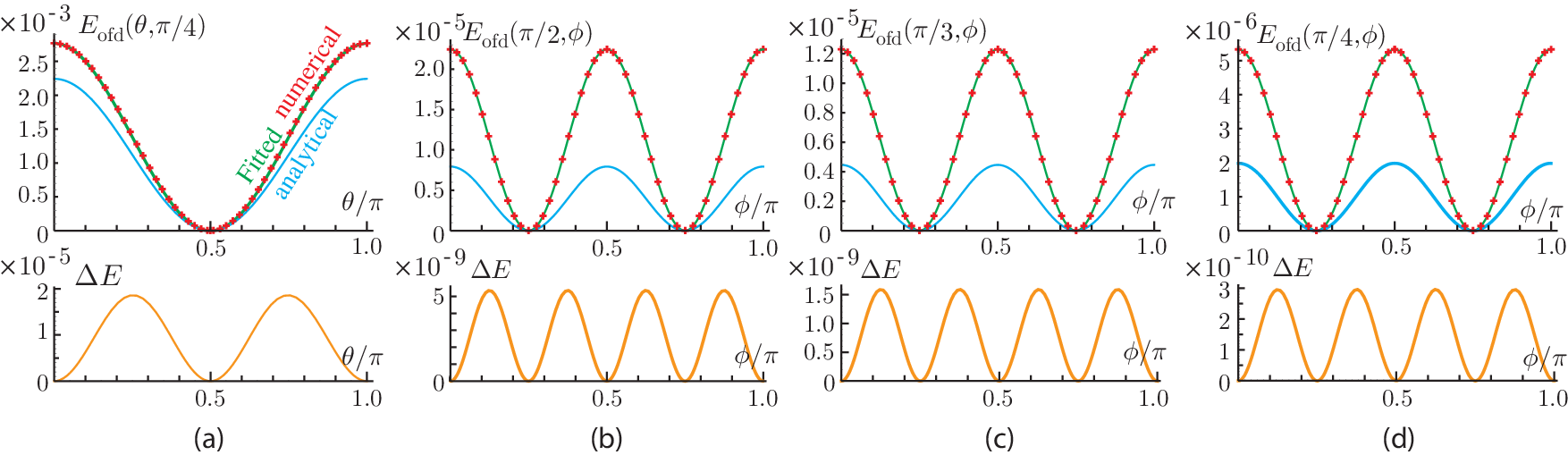}
    \caption{
	The same as Fig.S2 except using
	different parameters $t_\text{so}/t=1$, $n_0U/t_0=1$.
	The numerical results are represented by the red \textcolor{red}{+} symbol,
	the analytical results from the perturbation theory
	are represented by the blue lines,
	and fitting results are represented by the green lines
	( namely, treating coefficients $A$ and $B$ just as fitting parameters).
	The difference between the numerical results and the fitting results
	$\Delta E=E_\text{numerical}-E_\text{fit}$
	are listed below each sub-figure to show the form of Eq.\eqref{eq:pb}
	is still very accurate even at large $t_\text{so}/t=1$.}
\end{figure}

    In fact, the higher order terms such as $ (|\psi_1|^2-|\psi_2|^2)^4 $ and $ [(\psi_1^*\psi_2)^2+(\psi_1\psi_2^*)^2]^4 $
    in Eq.\eqref{eq:pb1}  can also be determined from Fig.S2.
    The fact that the period of $\Delta E$ is only half of $E$ can be cast into
	these high-order terms.
	For example, the 4th order perturbation also contain $\theta$-only dependent terms
	like $\cos2\theta$ and $\cos4\theta$, but they are proportional to $t_\text{so}^4$.
	The $\cos2\theta$ terms do not change the form of Eq.\eqref{eq:pb1}, they can be absorbed into the coefficient $ A $ anyway.
    But $\cos4\theta$ does. So after the subtraction, only small sub-leading terms survive,
    then $\Delta E(\theta,\pi/4)\sim \cos4\theta$ can be cast into the form $ C (|\psi_1|^2-|\psi_2|^2)^4 $.
	Similar arguments hold also for $\phi$ and lead to$\Delta E(\pi/2,\phi)\sim \cos8\phi$ which can be cast into
    $ D [(\psi_1^*\psi_2)^2+(\psi_1\psi_2^*)^2]^4 $. Both $ C $ and $ D $ are found to be positive.


For the local property, we focus on the
derivative with respect to $\theta$ and $\phi$:
\begin{align}
    A'(\theta_0)
	=\frac{\partial^2}{\partial \theta^2}
	E_\text{ofd}(\theta,\phi)\Big|_{\theta=\theta_0,\phi=\pi/4}\>,
    \qquad
    B'(\theta_0)
	=\frac{\partial^2}{\partial \phi^2}
	E_\text{ofd}(\theta,\phi)\Big|_{\theta=\theta_0,\phi=\pi/4}
\end{align}
Analytical result Eq.\eqref{eq:pb} predicts
$A'_{an}=-A\cos2\theta_0$ and $B'_{an}=B\sin^4\theta_0$,
and numerical derivatives can be evaluated from the finite differences of Eq.M3.
We plot both results in Fig.S3(a) and (b),
and find the differences are quite small.
The insert of Fig.S3(b) also confirms  $B'\sim \theta^4$ behaviour.

As a final comparison,
we evaluate $A'$ and $B'$ as a function of $t_\text{so}$
and plot them in Fig.S3(c) and (d).
When extrapolating the data to the $t_\text{so}\to0$ limit,
they reach agreement with the perturbation results with a high accuracy.
The insert of Fig.S3 (c) and (d) also shows that
the leading behaviours of $A'\sim t_\text{so}^2$ and $B'\sim t_\text{so}^4$,
in consistent with the perturbation theory listed below Eq.{eq:pb}.

In conclusion,
we show that the $E_\text{ofd}$ obtained from the perturbation theory
fit very well the data from the direct numerical evaluations, even upto $t_\text{so}/t=1$.

\begin{figure}[!htb]
\centering
    \includegraphics[width=\linewidth]{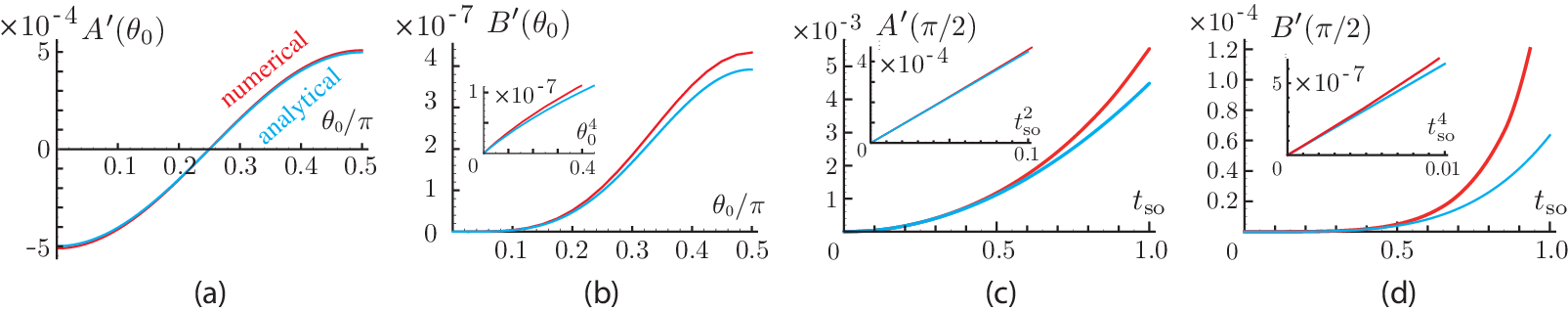}
    \caption{(a) Coefficient $A'$ as a function of $\theta_0$,
    (b) Coefficient $B'$ as a function of $\theta_0$,
    The insert shows $B'\sim\theta_0^4$ when $\theta_0$ is small.
    (c) Coefficient $A'$ as a function of $t_\text{so}$,
    The insert shows $A'\sim t_\text{so}^2$ when $t_\text{so}$ is small. 	
    (d) Coefficient $B'$ as a function of $t_\text{so}$.
    The insert shows $B'\sim t_\text{so}^4$ when $t_\text{so}$ is small.}
\end{figure}

\section{C. Some additional calculations using the effective GL action }

 In this appendix, we will use the effective action Eq.\ref{blqh} to perform alternative calculations to
 compute the excitation spectrum in both $ h>h_c $ and $ h < h_c $. They reproduce the results Eq.\ref{highpm} and
 Eq.\ref{lowpm} respectively achieved in the main text and also provide additional physical insights into the problem.

\subsection{1. A rotated  polar coordinate calculation at $h>h_c$}
When $h>h_c$, the system is in the Z-FM superfluid phase
with the saddle point solution Eq.\ref{zfmsad} $\bar{\rho}_1=\rho_0=-r/U$, $\bar{\rho}_2=0$, and no constraints on $\bar{\theta}_1$ and $\bar{\theta}_2$.
In order to avoid any singularities in using the polar coordinates, we introduce two rotated order parameters:
\begin{equation}
\psi'_1=(\psi_1+\psi_2)/\sqrt{2},~~~\psi'_2=(\psi_1-\psi_2)/\sqrt{2}
\end{equation}
then parameterize them as
$\psi'_1=\sqrt{\rho'_1}e^{i\theta'_1}$ and $\psi'_2=\sqrt{\rho'_2}e^{i\theta'_2}$.
It is easy to identify the saddle point solution
$\bar{\rho}_1=\rho_0, \bar{\rho}_2=0, \bar{\theta}_1=\bar{\theta}_2=0$ in old coordinates system
correspond to $\bar{\rho}'_1=\bar{\rho}'_2=\rho_0/2, \bar{\theta}'_1=\bar{\theta}'_2=0$ in new coordinate system.

For the notation simplicity, in the following, we drop the $'$ in $\rho$ and $\theta$.
After including fluctuations as
$\psi_\alpha=\sqrt{\bar{\rho}_\alpha+\delta \rho_\alpha}e^{i(\bar{\theta}_\alpha+\delta\theta_\alpha)}$,
then
\begin{align}
    \mathcal{L}_\text{ZF}
	=&i\delta \rho_1\partial_\tau\delta\theta_1+i\delta \rho_2\partial_\tau\delta\theta_2
	+\frac{v^2}{2\rho_0}[(\nabla\delta \rho_1)^2+(\nabla\delta \rho_2)^2]
	+\frac{1}{2}v^2\rho_0[(\nabla\delta \theta_1)^2+(\nabla\delta \theta_2)^2]\nonumber\\
	&+\frac{U}{2}(\delta \rho_1+\delta \rho_2)^2
	+\frac{A}{2}[4\delta \rho_1\delta \rho_2-\rho_0^2(\delta\theta_1-\delta\theta_2)^2]
	+\frac{h}{2}[\frac{1}{\rho_0}(\delta \rho_1-\delta \rho_2)^2+\rho_0(\delta\theta_1-\delta\theta_2)^2]
\label{GL2}
\end{align}
where the linear term drops out due to the saddle point condition
$[r+(A+U)\rho_0-h](\delta\rho_1+\delta\rho_2)=0$.
From Eq.\eqref{GL2}, one can easily extract two eigen-modes in the long wavelength limit:
\begin{align}
	\omega_-&=\sqrt{2\rho_0(A+U)v^2k^2}\\
	\omega_+&=2(h-A\rho_0)+v^2k^2
\end{align}
  which are identical to Eq.\ref{highpm}.

 Similar to Eq.\ref{actionpm}, one can also introduce $ \pm $ in the new coordinates system:
  $\delta\rho_\pm=\delta \rho_{1}\pm \delta\rho_{2}$
and $\delta\theta_\pm=\delta\theta_{1}\pm\delta\theta_{2}$,
then
\begin{align}
    \mathcal{L}_\text{FM}
	&=\frac{i}{2}\delta \rho_+\partial_\tau\delta\theta_+
	+\frac{v^2}{4\rho_0}[(\nabla\delta \rho_+)^2+\rho_0^2(\nabla\delta\theta_+)^2]
	+\frac{A+U}{2}(\delta \rho_+)^2\nonumber\\
	&+\frac{i}{2}\delta \rho_-\partial_\tau\delta\theta_-
	+\frac{v^2}{4\rho_0}[(\nabla\delta n_-)^2+\rho_0[(\nabla\delta\theta_-)^2]
	+\frac{h-A\rho_0}{2\rho_0}[(\delta \rho_-)^2+\rho_0^2(\delta\theta_-)^2]	
\end{align}
which lead to exactly the same two eigen-modes.

\subsection{2. A mixed coordinate calculation at $h<h_c$}
One may also use the mixed coordinate system representation to calculate the excitation spectrum
in the CAFM SF at $h<h_c$. Namely, using the saddle point in Eq.\ref{canted} and substituting  $\psi_1=\sqrt{\bar{\rho}_1+\delta\rho_1}e^{i(\bar{\theta}_1+\delta\theta_1)}$
and $\psi_2=\sqrt{\bar{\rho}_2}e^{i\bar{\theta}_2}+\delta\psi_2$ into Eq.\ref{blqh} leads to
the quadratic part of the Lagrangian:
\begin{align}
    \mathcal{L}_\text{CA}
	=&i\delta\rho_1\partial_\tau\delta\theta_1
	+\delta\psi_2^*\partial_\tau\delta\psi_2
	+\frac{v^2}{4\bar{\rho}_1}(\nabla\delta\rho_1)^2
	+v^2\bar{\rho}_1(\nabla\delta\theta_1)^2
	+v^2|\nabla\delta\psi_2|^2+(r+h)|\delta\psi_2|^2	\nonumber\\
	&+\frac{1}{2}(U+A)(\delta\rho_1)^2+8B\rho_1^2\rho_2^2(\delta\theta_1)^2
	+[2(U+A)\rho_2+(U-A)\rho_1+4B\rho_1^2\rho_2]|\delta\psi_2|^2	\nonumber\\
	&-i[\frac{1}{2}(U+A)-2B\rho_1^2]\rho_2(\delta\psi_2)^2
	+i[\frac{1}{2}(U+A)-2B\rho_1^2]\rho_2(\delta\psi_2^*)^2		\nonumber\\
	&+(U-A)\rho_2^{1/2}\delta\rho_1(e^{-i\pi/4}\delta\psi_2+e^{+i\pi/4}\delta\psi_2^*)
	+8B\rho_1^2\rho_2^{3/2}\delta\theta_1(e^{+i\pi/4}\delta\psi_2+e^{-i\pi/4}\delta\psi_2^*)
\label{GL3-2}
\end{align}
From Eq.\eqref{GL3-2}, one can extract the two eigen modes
which are exactly the same as Eq.\ref{lowpm} obtained in the polar coordinates in the main text.


\begin{thebibliography}{99}
\bibitem{origin}  The end of the boundary sits at the origin $t_\text{so}=0, h=0$ which
	has the exact SU(2) symmetry, the other parts only have the emergent U(1) symmetry.
	The whole line has the dynamic exponent $z=2$. As demeonstrated in the text,
    it has the same universality class as the zero density SF-Mott transition subject to a new
    marginally and dengerously irrelevant anisotropic term.
\bibitem{RBZ} This is because there are 2 valleies in Eq.\ref{deg}, 2 spins and the particle-hole spaces leading to the $ 8 \times 8 $
Bogoliubov transformation and the Reduced BZ in Eq.\ref{ofdleft}.
\bibitem{RGflow}   For notational simplicity, dropping the subscript $ 2 $ and writing $ \psi $ in terms of its real and imaginary part
   $ \psi=\phi_1+ i \phi_2 $, then the interaction terms in Eq.\ref{critical2} can be written as
   $ (u-6\gamma) ( \phi^2_1 + \phi^2_2)^2 + 8 \gamma ( \phi^4_1 + \phi^4_2) $ where the second term
   is the cubic anisotropic term with $ N=2 $ real components. Its RG flow at $ \epsilon=3-d > 0 $ was well known at any $ N $ component
   with 4 RG fixed points (1) Gaussian fixed point
   $ u^{*}=v^{*}=0 $ (2) Ising fixed point $ u^{*}=0, v^{*} > 0 $ (3) $ O(N) $ fixed point $ u^{*} > 0, v^{*} = 0 $
   and (4)  Cubic fixed point $ u^{*} > 0, v^{*} \sim (N-4) \epsilon  \neq 0 $.
   The fixed point structure changes from $ N > 4 $ where the cubic fixed point $ u^{*} > 0, v^{*} > 0 $ is the only stable fixed point
   to $ N < 4 $ where the $ O(N) $ fixed point is the only stable fixed point.
   So when $ N=2 < 4 $, it belongs to the latter case where the cubic anisotropy is irrelevant near the 3d XY fixed point.
   At the upper critical dimension $ \epsilon =0 $, all the 4 fixed points  converge to the  Gaussian fixed point.
   It is interesting to repeat this calculation at $ d=2, z=2 $ where the upper critical dimension $ d=2 $ and $ \epsilon=0 $.
   For example, we will  put $ u_2 $ and $ \gamma_1 $ in Eq.\ref{critical2} at the same footing to perform the RG analysis, also evaluate the new
   logarithmic corrections to Eq.\ref{tranMag} due to the two marginally irrelevant couplings.
   In fact, just like the transverse magnitization in Eq.\ref{tranMag}, the specific heat $ C_R $, the Lyapunov
   exponent $ \lambda_L $, the butterfly velocity $ v_{B,R} $ will all receive a logarithmic factor at a finite $ T $.
\bibitem{anotherwriting} More straightforwardly and intuitively, just like in BLQH \cite{blqh3}, one can write the interaction term as
 $ \frac{U}{2}[ |\psi_1|^2+|\psi_2|^2 - \rho_0 ]^2 $ where $ \rho_0$ is the average total atom density
 ( or electron density  in BLQH ). Expanding it out automatically leads to $ r=-\rho_0 U $.
\bibitem{blqhdrop}  In the BLQH \cite{blqh1,blqh2,blqh3}, the two terms in the gradient of  $ \delta\rho_+ $ and $ \delta\rho_- $ were dropped due to
their subleading to the corresponding mass terms.
\bibitem{pmmode} In the absence of Zeeman field $ h=0 $, the $ \pm $ modes in Eq.\ref{actionpm} are decoupled,
so the $ \pm $ mode stand for the gapless SF mode and gapped roton mode in $ \pm $ channels respectively.
However, when $ h > 0 $, there is a mixing between the two channels whci lead to two new eigen-modes which are neither
$ + $ or $ - $ mode.  But we still use the same symbols.
\bibitem{dimen} In the whole manuscript, we take the lattice constant $ a=1 $, so $ ka $ and $ x/a $ are  dimensionless, then
the butterfly velocity $ v_B $ carry the dimension of energy.
\end{thebibliography}
\begin{thebibliography}{99}
\bibitem{more} To calculate the gap at $ h=0 $, this expansion of the effective potential near the minimum $ \theta=\pi/2, \phi=\pi/4 $ is enough.
However, to get the correct roton spectrum at any $ h >0 $  at any $ k $ in the long wavelength limit,
one need to find the complete form of the the effective potential  in all ranges of  $ 0< \theta < \pi, 0< \phi < 2 \pi  $ which is found to be Eq.M5 and re-written as Eq.M6 in the main text. Most importantly,
as shown in the main text, just from the symmetry point of view, the $ A $ term and $ B $ term
in the phenomelogical GL action Eq.\ref{blqh} precisely match Eq.M6 at the weak coupling limit. This endows the effective potential generated by OFQD
from a new and profound perspective.
\end{thebibliography}
\end{document}